\documentclass[journal]{IEEEtran}

\usepackage{amsmath}
\usepackage{bm}
\usepackage{graphicx}
\usepackage[ruled,linesnumbered]{algorithm2e}
\usepackage{algpseudocode}
\usepackage{amsfonts,amssymb}
\usepackage{mathrsfs}
\usepackage{multirow}
\usepackage{fmtcount}
\usepackage{url}
\usepackage{setspace}
\usepackage{blindtext}
\usepackage{xcolor,colortbl}
\usepackage{fancyhdr}
\usepackage{booktabs}
\usepackage{dblfloatfix}
\usepackage{anyfontsize}
\usepackage{enumerate}
\usepackage{ntheorem}
\usepackage{pifont}
\usepackage{hhline}
\usepackage{booktabs}
\usepackage{flushend}
\usepackage[flushleft]{threeparttable}
\usepackage[framemethod=default]{mdframed}
\usepackage{showexpl}

\usepackage{soul}
\DeclareRobustCommand\comment[1]{}
\DeclareRobustCommand\jiaqi[1]{}

\begin{document}

\title{\fontsize{23}{30}\selectfont Computing-Aware Routing for LEO Satellite Networks: A Transmission and Computation Integration Approach}

\author{Jiaqi Cao,\ 
Shengli Zhang,~\IEEEmembership{Senior Member,~IEEE,}\\
Qingxia Chen,\ 
Houtian Wang,\ 
Mingzhe Wang,\ 
Naijin Liu\ 

\thanks{Naijin Liu is the corresponding author.}      
\thanks{Jiaqi Cao and Shengli Zhang are with Shenzhen University, Shenzhen, 518052, P.R. China (e-mail: jiaqicao@szu.edu.cn; zsl@szu.edu.cn).}
\thanks{Qingxia Chen, Houtian Wang and Naijin Liu are with Qian Xuesen Laboratory of Space Technology, China Academy of Space Technology (e-mail: chenqingxia@qxslab.cn; wanghoutian@qxslab.cn; liunaijin@qxslab.cn).}
\thanks{Mingzhe Wang is with Tsinghua University (e-mail: wmzhere@gmail.com).}
}
\maketitle
\thispagestyle{fancy}

\begin{abstract}
The advancements of remote sensing (RS) pose increasingly high demands on computation and transmission resources.
Conventional ground-offloading techniques, which transmit large amounts of raw data to the ground, suffer from poor satellite-to-ground link quality.
In addition, existing satellite-offloading techniques, which offload computational tasks to low earth orbit (LEO) satellites located within the visible range of RS satellites for processing, cannot leverage the full computing capability of the network because the computational resources of visible LEO satellites are limited. This situation is even worse in hotspot areas.
	
In this paper, for efficient offloading via LEO satellite networks, we propose a novel computing-aware routing scheme. 
It fuses the transmission and computation processes and optimizes the overall delay of both.
Specifically, we first model the LEO satellite network as a snapshot-free dynamic network, whose nodes and edges both have time-varying weights. By utilizing time-varying network parameters to characterize the network dynamics, the proposed method establishes a continuous-time model which scales well on large networks and improves the accuracy.
Next, we propose a computing-aware routing scheme following the model. It processes tasks during the routing process instead of offloading raw data to ground stations, reducing the overall delay and avoiding network congestion consequently.
Finally, we formulate the computing-aware routing problem in the dynamic network as a combination of multiple dynamic single source shortest path (DSSSP) problems and propose a genetic algorithm (GA) based method to approximate the results in a reasonable time.
Simulation results show that the computing-aware routing scheme decreases the overall delay by 78.31\% compared with offloading raw data to the ground to process when the computing capability is 100 Giga floating-point operations per second (GFLOPS) which is a trivial computing capability supported by most LEO satellites.
	
\end{abstract}

\begin{IEEEkeywords}
LEO satellite network, remote sensing, computing-aware routing, dynamic network, genetic algorithm. 
\end{IEEEkeywords}
\IEEEpeerreviewmaketitle%

\section{Introduction}\label{Introduction}
\IEEEPARstart{R}{emote} sensing (RS) is playing an increasingly important role in Earth science, space science, and exploration science, such as environmental studies, military applications, hazard tracking and monitoring~\cite{plaza2007high}. 
To complete such space missions, resources in both computation and data transmission are needed.
On the one hand, the advancements in image processing and target recognition techniques, especially the application of machine learning techniques~\cite{7486259}, have led to a rapid increase in computational requirements~\cite{5678816,917889}. 
On the other hand, sensing technology improvements, such as hyperspectral image (HSI)~\cite{8113122}, enable increased data precision at the cost of a huge data volume of remote sensing images.
Conventionally, these images are offloaded to ground servers for computation.

In the aforementioned ground-offloading approach, 
satellites act as bent pipes to route massive raw data of RS tasks to ground servers for processing. 
While the ground servers are powerful in computing capability, the overall delay\footnote{The overall delay refers to the moment from a task is generated to the moment when the destination obtains the computing results of that task. For a computation task, \textit{both} transmission and computation processes affect the overall delay.} of the ground-offloading approach is still hard to meet the requirements of most RS applications which require real-time or near real-time processing capabilities~\cite{plaza2009special}. 
It is because transmitting raw data can be a significant bottleneck: perturbed by the atmosphere frequently~\cite{alonso2004performance}, the satellite-to-ground link (SGL) can be as low as 20 Mbps in state-of-the-art satellites~\cite{yost2021state}. To overcome the limitations of the ground offloading scheme, researchers have set their sights on onboard computing.

One promising target to offload computational tasks is low earth orbit (LEO) satellites.
Thanks to the development of LEO satellite computation capabilities, high-performance onboard computing provided by a large number of LEO satellites could alleviate these challenges by processing data before transmitting them to ground servers~\cite{2017Comparative}. This scheme can usually achieve impressive performance compared with ground offloading for the following reasons.
First, LEO satellites are geographically closer to RS satellites than ground servers, avoiding transmitting large amounts of raw data on SGLs whose data rate are relatively low.
Second, onboard processing significantly reduces the resulting data's volume (down to a few bits sometimes), which lowers the transmission delay in turn.
Despite the promising future of LEO satellite networks based offloading, major challenges need to be resolved first.

\textit{Challenge 1: Uneven Available-Resource Distribution}.
More than three quarters of the Earth's surface is covered by oceans and glaciers with no frequent human activity; instead, a large number of tasks generated by human activity are located on land, especially in hot spots (such as cities and ports).
As the distribution of offloaded tasks are unbalanced, available resources including computing and spectrum resources are also unevenly distributed on the LEO satellite network~\cite{6029964,8398222}. 
In this condition, LEO satellites nearby may not have sufficient computational resources.
As a result, computational tasks generated by remote sensing satellites need to be routed to farther satellites with sufficient computational resources for processing.
Frequently, the distance can be long enough so that multiple hops are required to reach LEO satellites. Therefore, a routing strategy is needed to transmit the tasks generated in hot spots to remote LEO satellites for computing.

\textit{Challenge 2: Inapplicable Terrestrial Routing Strategy}.
Existing terrestrial network routing strategies focus on static networks and commonly employ shortest path algorithms such as Dijkstra to find viable paths. 
However, such algorithms cannot be applied to LEO satellite networks because contemporaneous paths sometimes do not exist between the source and destination~\cite{madni2020dtn}. The phenomenon can be explained by the physical nature of LEO satellites: the high-speed relative motions between adjacent-orbiting satellites combined with the limited communication distance cause intermittent connectivity in LEO satellite networks.
Consequently, a routing strategy specific to LEO satellite networks is needed to accommodate its innate physical characteristics.


\textit{Challenge 3: Unaccounted Computation Cost}.
Previous routing strategies for LEO satellite networks focus on data transmission only. However, the transmission process and the computation process jointly determine the overall delay (i.e., the time from task generation to the arrival of computation results at the destination) of a computational task.
Existing routing strategies can only find the shortest path from a remote sensing satellite to a given LEO satellite used for computing, but cannot evaluate and select appropriate LEO satellites to conduct computation automatically.
Therefore, the satellite to be used for computation is still unanswered.

\begin{figure}[t]
	\centering
	\setlength{\abovecaptionskip}{-0.cm}
	\setlength{\belowcaptionskip}{-0.cm}
	\includegraphics[width=0.45\textwidth]{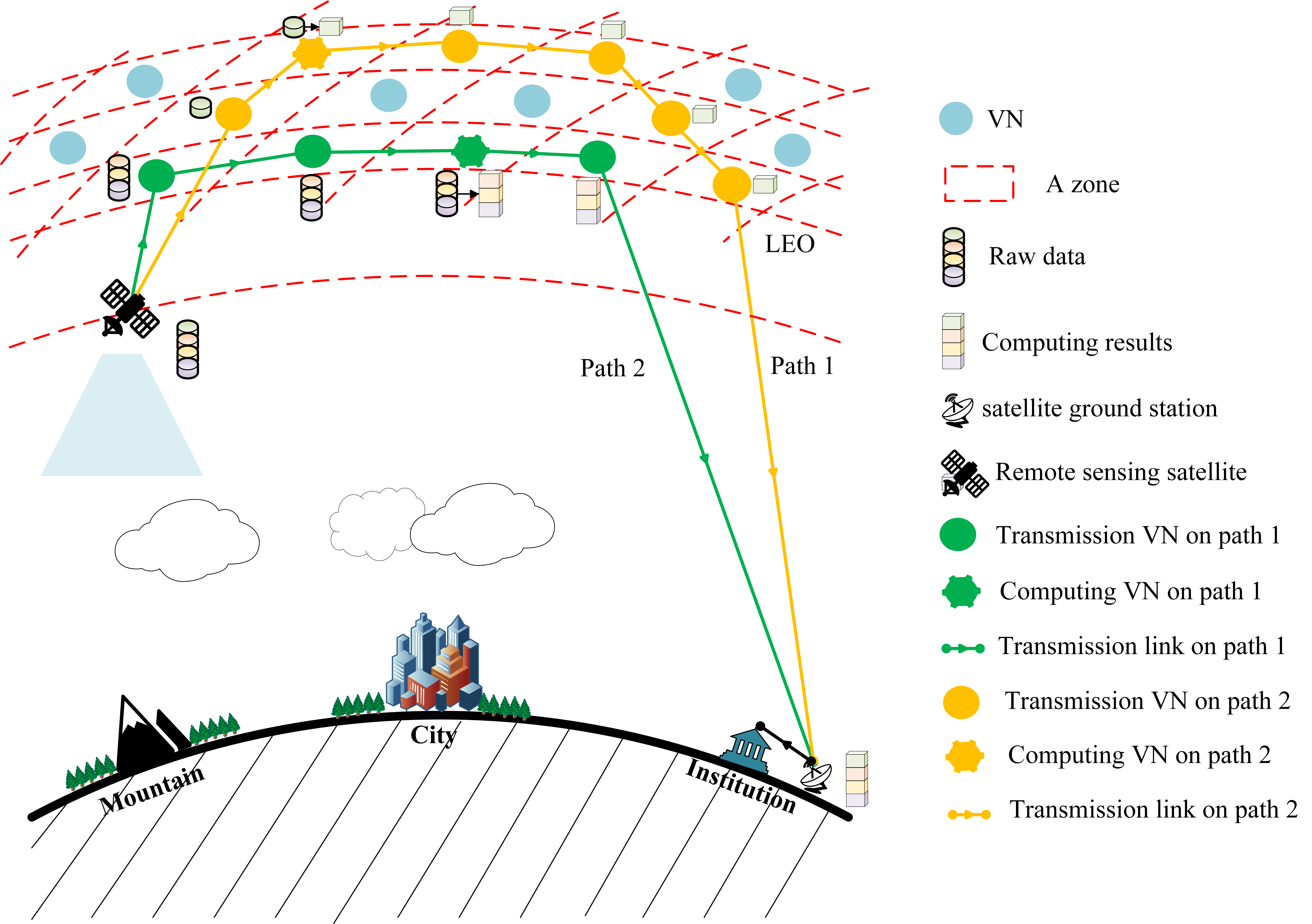}\\
	\begin{center}
		\caption{Example application of the computing-aware routing problem. Remote sensing satellites continuously capture hyperspectral images of the surveillance area. The institution on the ground needs to obtain the analysis results of these images. scheme.}\label{MobileStationaryDestinationRouting}
	\end{center}	
    \vspace{-1.0em} 
\end{figure}

\textit{Our Solution}.
For efficient task offloading via LEO satellite networks, we propose a novel computing-aware routing scheme to minimize the overall delay. As shown in Fig.~\ref{MobileStationaryDestinationRouting}, the computing-aware routing problem contains three sub-processes: route raw data of tasks from the source to the selected computing node; process tasks and generate computing results at the computing node; route the computing results from the computing node to the destination. Since the above transmission and computation processes affect the overall delay collaboratively, these processes are jointly optimized in the proposed computing-aware routing scheme.
For challenge 1, the introduction of routing extends the task offloading targets, so that satellites beyond the visible range can be scheduled for computation.
For challenge 2, the scheme models the network as a dynamic system, thus the proposed routing algorithm can tolerate the fast change of available resources and network topology by design.
For challenge 3, the joint optimization algorithm minimizes the overall delay of both the transmission and computation stages; in other words, it additionally considers the computation delay compared to existing routing algorithms.

\begin{table*}[b]
	\centering
	\vspace{-1.0em}
	\begin{spacing}{1}
		\caption{Comparison of Graph-Based Satellite Network Routing}\label{tab:GraphComparison}
	\end{spacing}
	\small
	\resizebox{0.95\textwidth}{!}{
	\begin{tabular}{ccccccccc}
		\toprule
		\multirow{2}{*}{Reference} & \multirow{2}{*}{\multirow{2}{*}{\begin{tabular}[c]{@{}c@{}}Network \\Modeling\end{tabular}}} & \multirow{2}{*}{Classification} & \multicolumn{2}{c}{Dynamics Representation} & \multicolumn{3}{c}{Graph/Model Size} & \multirow{2}{*}{Main Shortcomings}\\
		\cmidrule(lr){4-5} \cmidrule(lr){6-8}
		~ & ~&~ & Topology & Resource & Snapshot Number & Node Number & Edge Number & ~\\
		\midrule
		~\cite{el2015routing}& TG&\multirow{3}{*}{\begin{tabular}[c]{@{}c@{}}Snapshot\\~model\end{tabular}}  & \checkmark & & $ \propto N $ & $ O(|\mathbf{S}|) $ & $ O(|\mathbf{S}|) $ & \multirow{3}{*}{\begin{tabular}[c]{@{}c@{}}Isolated snapshots split the connectivity of the \\whole network and consume massive memory\end{tabular}} \\
		~\cite{9524989}& VT&~  & \checkmark & & $ \bowtie |\mathbf{S}| $ & $ O(|\mathbf{S}|) $ & $ O(|\mathbf{S}|) $ & ~\\
		~\cite{6412688}& VN&~  & \checkmark & & $ \bowtie (|\mathbf{S}|,|\mathbf{Z}|) $ & $ O(|\mathbf{Z}|) $ & $ O(|\mathbf{Z}|) $ & ~ \\
		\midrule
		~\cite{burleigh2011contact}& CG&\multirow{4}{*}{\begin{tabular}[c]{@{}c@{}}Non-snapshot\\model\end{tabular}}  & \checkmark & & N/A & $ O(|\mathbf{S}|) $ & $ O(|\mathbf{S}|) $ & No assurance for task demands and low resource utilization \\
		~\cite{9003306,9067004}& TEG &~  & \checkmark & \checkmark & N/A & $ O(N\times |\mathbf{S}|) $ & $ O(N\times |\mathbf{S}|) $ & High storage overhead and computational complexity \\
		\cite{kohler2002time,8766128}& TAG/STAG&~  & \checkmark & \checkmark & N/A & $ O(|\mathbf{S}|) $ & $ O(|\mathbf{S}|^2) $ & Excessive model size for highly dynamic networks \\
		Our work& SFDNM&~  & \checkmark & \checkmark & N/A & $ O(|\mathbf{Z}|) $ & $ O(|\mathbf{Z}|) $ & ~     \\
		\bottomrule
	\end{tabular}
	
}
	
	\tiny
	\begin{tablenotes}
		\item $ \propto $: Proportional to, $\bowtie$: related to.
		\item $ |\mathbf{S}| $: Satellite number, $ |\mathbf{Z}| $: zone (i.e., VN) number, $ N $: slot number. 
	\end{tablenotes}
\end{table*}

\subsection{Main Contributions}\label{MainContributions}
\subsubsection{A Snapshot-Free Dynamic Network Model} We propose a snapshot-free dynamic network modeling method for LEO satellite networks for cross-time pathfinding with low memory consumption. It can represent both resource dynamics and topology dynamics. It utilizes time-varying edge weights and node weights to represent the \textit{resource dynamics} related to the transmission and computation processes, respectively. In addition, instead of shielding the dynamics, the proposed model converts the \textit{topology dynamics} into the association dynamics between satellites and virtual nodes (VNs), which represents self-loops, special edge and node weights.

\subsubsection{A Computing-Aware Routing Scheme} We propose a novel routing scheme for LEO satellite networks, which processes tasks during the routing process. By performing onboard computing, the proposed scheme can achieve significant bandwidth savings especially for satellite-ground links.
Because the slow process of offloading raw data to ground servers is avoided, the proposed routing scheme can reduce the overall delay.
This scheme optimizes the computation and transmission processes jointly, therefore, tasks can be offloaded to the optimal satellite via the optimal path.
Furthermore, any satellite with sufficient resources can be selected as the offloading target. Since LEO constellations, especially for giant LEO constellations, usually consist of a large number of satellites, a large amount of computing resources could be utilized for computing-aware routing.

\subsubsection{A Genetic Algorithm Based Approximation Method} Since the LEO satellite network is highly dynamic and the on-board resources (related to node weights) need to be considered in the computing-aware routing, we formulate the LEO satellite network computing-aware routing problem based on the proposed edge-weighted and node-weighted dynamic network and convert it to a set of dynamic single source shortest path (DSSSP) problems. Due to the dynamics in graphs, conventional shortest path algorithms such as the static Dijkstra algorithm cannot be used to solve the problem; thus, we propose a genetic algorithm (GA) based method to approximate the results in reasonable time.


The rest of this paper is organized as follows. 
In Section~\ref{RelatedWork}, the related studies are summarized. 
Section~\ref{LEOCharacteristics} investigates the state-of-the-art computing and transmission capabilities of LEO satellites.   
Section~\ref{SystemModel} presents the network model, traffic model, and delay model adopted in this paper.  
Section~\ref{LEOCharacteristics} proposed a snapshot-free dynamic network modeling method. 
The computing-aware routing problem is formulated in Section~\ref{ProblemFormulation}. In Section~\ref{AlgorihmDesign}, a GA-based path finding algorithm is proposed. Simulation results and analyses are given in Section~\ref{Simulations}. Conclusions are drawn in Section~\ref{Conclusions}.

\section{Related Work}\label{RelatedWork}
Due to the superiority in latency, cost, development cycle, etc., the LEO satellite network is deemed as the most prospective satellite mobile communication system. Therefore, a lot of studies on LEO satellites and LEO satellite networks have been conducted in academia and industry. Several aspects relevant to this paper are introduced below.

\subsection{Routing Strategies for LEO Satellite Networks}\label{RelatedWork-Routing}
The ever changing relative positions of satellites bring constant changes in network topology; even worse, for specific instants, no contemporaneous path exists between the source and destination nodes~\cite{madni2020dtn}. Therefore, routing is a challenging issue in LEO satellite networks. 

Graph theory is an effective mathematical tool to model the network, providing a basis for the routing design~\cite{DBLP:conf/mobicom/WhitbeckACG12}. 
Therefore, graph-based routing strategies for LEO satellite networks have attracted widespread attention, which are summarized in Table~\ref{tab:GraphComparison}.
\textit{Contact graph routing (CGR)} was proposed for dynamic routing over the time-varying topology of satellite networks by NASA~\cite{burleigh2011contact}.
The basic idea of CGR is to utilize a scheduled contact graph for pathfinding.
The \textit{contact graph (CG)} records the network dynamics with a ``contact plan'' which is a time-ordered list of scheduled changes of network topology~\cite{7060480}.
However, they could not ensure the mission’s demands and could not fully utilize the resources~\cite{8766128}.

To deal with the dynamics in satellite networks, some existing works proposed snapshot-based network modeling methods, such as the \textit{temporal graph (TG)}, the \textit{virtual topology (VT)} model and the \textit{VN} model, and the corresponding routing strategies.
The basic idea of TG~\cite{el2015routing} is to divide the system period into a set of discrete slots. In this way, the dynamic network can be represented as a set of static topology graphs. 
Similarly, VT-based network models~\cite{9524989,8352859,soret2019autonomous,7023604} are presented based on the determinism of satellite movements, representing a satellite network as a time-evolving and predictable network~\cite{2017Organized}. It considers a LEO satellite network as a discrete-time network, and assumes a fixed topology in each time interval~\cite{634801}, which is called a snapshot. Routes are defined at each snapshot by using this method. 
VN-based network models~\cite{6412688,4384138} are composed of different logical locations, which are static and disjoint zones of Earth (i.e.,\ latitude and longitude), associated with the nearest satellites~\cite{2001A}. 
The assignment between the logical locations and satellites changes due to satellite movements. With this architecture, each change on the satellite assignment represents a new snapshot~\cite{8332924}. Each snapshot could be considered as a mesh network presenting a static state of the network topology.
These methods split the dynamic network into multiple snapshots, where each snapshot corresponded to a constant network topology during a time slot~\cite{634801}.
Obviously, these methods only look for paths within a single snapshot and ignore the relationship between adjacent snapshots~\cite{9003306}. A snapshot is defunct if the duration of the generated route exceeds the snapshot's valid time. Transmitting tasks based on defunct routes may cause routing failures, especially for highly dynamic LEO satellite networks. 
Furthermore, as shown in Table\ref{tab:GraphComparison}, these snapshot-based models would consume massive memory resources because a prohibitively large number of snapshots would be generated as the time increases or the network expands~\cite{li2017maximum,8332924}.

To overcome the drawbacks of the snapshot-based routing strategies, the \textit{time-expanded graph (TEG)}~\cite{9003306,9067004,8719020,7341215,7510734} was proposed to establish connections between networks of adjacent slots. It duplicates the original network for each time slot and builds edges connecting each node and its copy at the next slot to represent the data storage.
Indeed, TEGs are essentially an expansion of static graphs, and hence many standard flow maximization algorithms can be applied to time-expanded graphs~\cite{li2017maximum}. 
Although the TEG significantly increases the connection of snapshots~\cite{7179361}, it also incurs high overhead in storage and algorithm~\cite{li2017maximum}. 

The \textit{time aggregated graph (TAG)}~\cite{kohler2002time,george2007spatio} aggregates the time-dependent attributes over edges and nodes. It represents the time-variance of attributes by modeling them as time series. 
To consider the buffer size constraint of each relay node in TAG, authors in~\cite{8766128,li2017maximum,zhang2017storage} proposed the \textit{storage time aggregated graph (STAG)}.
Although the TAG and STAG models could capture the possibility of edges and nodes being absent during certain instants of time~\cite{george2009time}, these methods still face the problem of edge explosion when modeling highly dynamic networks, such as giant LEO constellations. More specifically, when the periods (altitudes) of the adjacent orbits of the LEO constellation are different, any two satellites in adjacent orbits may establish inter-satellite link (ISL) within a certain period of time. In this condition, when constructing TAGs or STAGs, any two satellites in adjacent orbits should be connected with an edge. Consequently, the edge numbers in these graphs are quadratic to satellite numbers.

To overcome the shortcomings of these existing network modeling methods and routing strategies for the LEO satellite network, we propose a novel dynamic network model which can represent both resource dynamics and topology dynamics with low complexity and the corresponding computing-aware routing scheme.
\subsection{Computing and Transmission Joint Optimization}
Some existing works investigated the computation and transmission joint optimization for satellite networks. In these works, satellites are not connected with each other.
The work in~\cite{9383778} and~\cite{9515574} investigated the joint computation assignment and resource allocation problem in multi-tier computing architectures composed of mobile devices, LEO satellites, etc.
Authors in~\cite{9344666} and \cite{9793590} proposed hybrid computation offloading architectures to solve the joint computation and resource allocation problem, where computing tasks could be offloaded to both ground servers and visible LEO satellites. 

Some studies discussed the computation and transmission integration problem in terrestrial networks.
The work in~\cite{8387798} discussed the joint communication and computing resource allocation in a two-tier device--cloud network, where tasks could be processed locally, in the edge cloud, or both. 
Authors in~\cite{8663968} and~\cite{8713498} investigated the task offloading problem in fog-enabled cellular networks where radio, caching, and computing were jointly optimized.
The work in~\cite{9447257} and~\cite{8877759} proposed joint communication and computing resource scheduling approaches for unmanned aerial vehicle (UAV)-assisted local--edge/local--edge--cloud computing systems, where each UAV worked as an edge computing devices to assist devices within its communicable range.
Authors in~\cite{8936886} developed a cloud-fog-device computing architecture for internet of things (IoT), where the offloading ratio, transmission power,	 and local CPU computation speed were jointly optimized. 

Although the studies mentioned above jointly optimize the allocation of multiple resources, they still fail to achieve network-wide computation offloading. 
This is because the networks in the above studies are \textit{tree networks}, where tasks cannot be forwarded to other computing devices in the same network tier.
To overcome this limitation, LEO satellites in this paper are connected with ISLs, which form a \textit{mesh network}. 
In this condition, computing tasks can be offloaded to any satellite via routing, which makes it possible to extend the offloading targets to the entire network. However, the network-wide computation offloading for LEO satellites brings a novel challenge: the transmission path needs to be optimized as well. To address this challenge, we propose a computing-aware routing scheme to jointly optimize the resources and transmission paths. 

\section{State-of-the-Art LEO Satellite Capabilities}\label{LEOCharacteristics}
As the computational requirements and data volume of space missions increase, unprecedented interest and efforts have been devoted to enhancing the computing and transmission capabilities of LEO satellites. 
\subsection{Computation Capability of LEO Satellites}\label{LEOComputation}
For next-generation science and defense missions, spacecrafts such as LEO satellites must provide advanced processing capability to support a variety of computationally intensive tasks~\cite{geist2019spacecube}. 
The desire for even more onboard processing capacity has led to the development of onboard computing systems. The computing capabilities of some typical ;onboard computing systems are summarized in Table~\ref{Onboard Computing Systems}.

\begin{table}[h]
	\centering
	\vspace{-0.5em}
	\begin{spacing}{1}
		\caption{Computing Capabilities of Onboard Computing Systems (in GFLOPS)}\label{Onboard Computing Systems}
	\end{spacing}
	\small
	\resizebox{0.49\textwidth}{!}{
	\begin{tabular}{cccc} 
		\rowcolor[rgb]{0.812,0.835,0.918} Product             & Processor                                                                                                                  & \begin{tabular}[c]{@{}>{\cellcolor[rgb]{0.812,0.835,0.918}}c@{}}Computing\\Capability\end{tabular} & Reference                                     \\
		Xiphos
		
		Q7S                                           & Xilinx Zynq 7020                                                                                                           & 180                                                                                                & \cite{XiphosQ7S}                              \\
		\rowcolor[rgb]{0.914,0.922,0.961} Xiphos Q8S          & Xilinx Ultrascale+                                                                                                         & 1800                                                                                               & \cite{XiphosQ8S}                              \\
		BAE RAD5545                                           & RAD5545                                                                                                                    & 3.7                                                                                                & \cite{BAERAD5545}                             \\
		\rowcolor[rgb]{0.914,0.922,0.961} Innoflight CFC-500  & \begin{tabular}[c]{@{}>{\cellcolor[rgb]{0.914,0.922,0.961}}c@{}}Xilinx Kintex Ultrascale+, \\NVIDIA TK1\end{tabular}       & 1290                                                                                               & \cite{CFC500}                                 \\
		MOOG G-Series Steppe Eagle                            & AMD G-Series compatible                                                                                                    & 75                                                                                                 & \cite{MOOG}                                   \\
		\rowcolor[rgb]{0.914,0.922,0.961} MOOG V-Series Ryzen & AMD V-Series compatible                                                                                                    & 1000                                                                                               & \cite{MOOG}                                   \\
		Unibap iX5-100                                        & \begin{tabular}[c]{@{}c@{}}Microchip SmartFusion2,\\AMD G-Series SOC\end{tabular}                                          & 127                                                                                                & \cite{iX5-100,bruhn2020enabling}              \\
		\rowcolor[rgb]{0.914,0.922,0.961} Unibap iX10-100     & \begin{tabular}[c]{@{}>{\cellcolor[rgb]{0.914,0.922,0.961}}c@{}}Microchip PolarFire, \\AMD V1605b (Ryzen)\end{tabular}     & 3600                                                                                               & \cite{iX10-100,bruhn2020enabling}             \\
		SpaceCube v2.0                                        & Xilinx Virtex 5                                                                                                            & 200                                                                                                & \cite{SpaceCube2.0}                           \\
		\rowcolor[rgb]{0.914,0.922,0.961} SpaceCube v3.0      & \begin{tabular}[c]{@{}>{\cellcolor[rgb]{0.914,0.922,0.961}}c@{}}Xilinx Kintex UltraScale,\\ Xilinx Zynq MPSoC\end{tabular} & 590                                                                                                & \cite{geist2019spacecube,SpaceCube3.0Patent}  \\
	\end{tabular}
	}
\end{table}

It can be concluded from Table~\ref{Onboard Computing Systems} that existing onboard computing systems can provide thousands Giga floating-point operations per second (GFLOPS) of computing capability. For example, the national aeronautics and space administration (NASA) Goddard Space Flight Center (GSFC) developed SpaceCube v3.0 in 2019~\cite{9546282}. It contains a Xilinx Kintex UltraScale with a Xilinx Zynq MPSoC to provide 10--100x or more performance over other flight single-board computers~\cite{geist2019spacecube}. In specific, the computing capacities of these systems on chips (SoCs) both exceed 100 GFLOPS. 
Although the computing capability of LEO satellites is not yet comparable to that of geosynchronous equatorial orbit (GEO) satellites and ground servers, the prospect and importance of increasing the computing capability of LEO satellites has been recognized and a great deal of research has been invested, indicating a promising future for onboard computing.

\subsection{Data Rate of LEO Satellites}\label{LEODataRate}
ISLs in free space are usually higher in data rate. For instance, the data rate of optical ISLs can achieve 5 Gbps~\cite{del2019technical}. Mynaric's laser terminal for LEO constellations is capable of delivering 10 Gbps with a low SWaP unit over a wide range of constellation configurations~\cite{10.1117/12.2545629}. It can operate within densely packed constellations with intra/inter-plane link distances up to 7,800 km.

In contrast, the data rate of the satellite-to-ground link cannot keep up with the speed of the inter-satellite link due to the perturbation induced by the atmosphere~\cite{alonso2004performance}. The downlink data rate for state-of-the-art satellites ranges from 20 Mbps to 1 Gbps~\cite{yost2021state}. For example, the CubeSat lasercom module by Hyperion Technologies enables a bidirectional space-to-ground communication link between a CubeSat and an optical ground station, with a downlink speed up to 1 Gbps and an uplink data rate of 200 Kbps~\cite{yost2021state}. The limited SGL transmission capability further promotes the application of on-board computing.

\section{System Model}\label{SystemModel}
\subsection{Network Model}\label{NetworkProperties}
\begin{figure}[htbp]
	\centering
	\vspace{-1.0em}
	\setlength{\abovecaptionskip}{-0.cm}
	\setlength{\belowcaptionskip}{-0.cm}
	\includegraphics[width=0.48\textwidth]{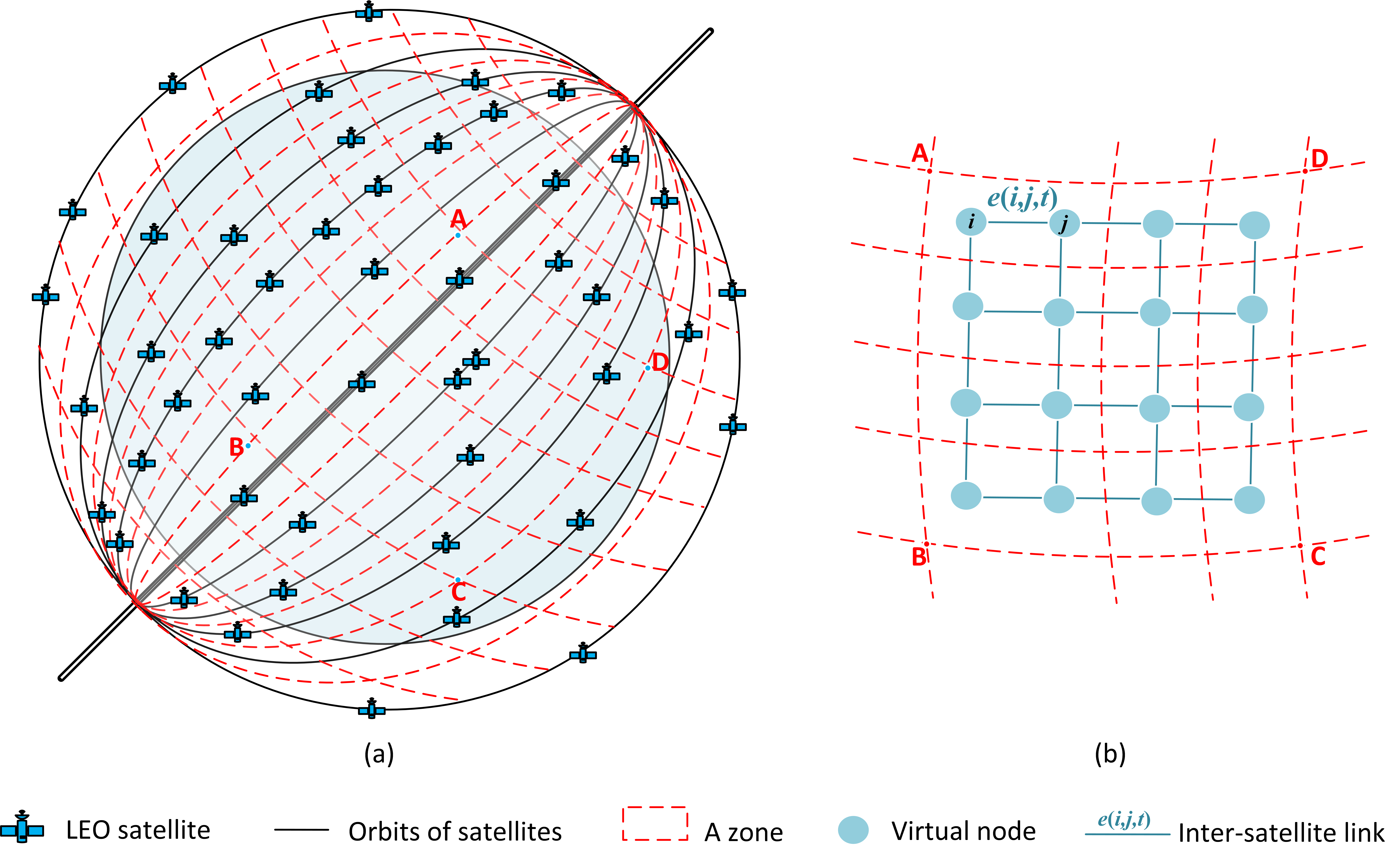}\\ 
	\begin{center}
		\caption{(a) LEO satellite network with various types of LEO satellites, (b) Snapshot-free dynamic network model.}\footnote{Type 1 and 2 in Fig 1 indicate that the proposed dynamic network model is compatible with LEO satellites having different properties, such as computing capability.}\label{VirtualSatelliteMeshNetwork}
	\end{center}
	\vspace{-1.0em}	 
\end{figure}

As shown in Fig.~\ref{VirtualSatelliteMeshNetwork} (a), the LEO satellites are uniformly distributed over orbits at an altitude of $ h $ kilometers. The satellites on the same orbit are uniformly distributed. The set of satellites is denoted by $ \textbf{S} = \{ S_1,S_2,\dots,S_{|\textbf{S}|}\} $ and the set of orbits is denoted by $ \textbf{O} = \{ O_1,O_2,\dots,O_{|\textbf{O}|}\} $. The orbit inclination $ i_0 $ determines the latitude of coverage. The orbital period is $ T_O = 2\times\pi\times\sqrt{{(R_e+h)}^3/(G\times M_e)} $, where $ R_e $ and $ M_e $ represent the radius and mass of the earth respectively, and $ G $ is the gravitational constant. 

In Fig.~\ref{VirtualSatelliteMeshNetwork} (a), the Earth is divided into multiple static disjoint zones according to longitude and latitude. These zones are stationary with respect to the ground.
Each static zone corresponds to a VN. A device in space (such as satellites) associated with a VN implies that its sub-satellite point (on the ground) is located in the zone corresponding to the VN. The association between satellites and VNs is changing over time. In this way, the \textit{topology dynamics} of the LEO satellite network are converted into the \textit{association dynamics} between satellites and VNs (addressed in Section~\ref{Association}).

Fig.~\ref{VirtualSatelliteMeshNetwork} (b) shows a part of the VN network shown in Fig.~\ref{VirtualSatelliteMeshNetwork} (a). The edges between each pair of VNs represent the communication links between the associated satellites. The inter-satellite and satellite-ground connection strategies are stated as follows.
\begin{itemize}
	\item \textbf{Inter-satellite connections}: each satellite has four inter-satellite links with its neighbors where two are intra-plane and two are inter-plane. 
	\item \textbf{Satellite-ground connections}: a satellite can communicate with a ground station only when the elevation angle between them is greater than the minimum elevation angle. For simplicity, in this paper we assume that a satellite can communicate with ground stations in a zone when its sub-satellite point is located in that zone.
\end{itemize}

In this VN network, the transmission resources are related to edges; whereas, the computing resources are related to VNs. The available resources decrease as they are occupied and increase as they are released, which forms \textit{resource dynamics} (addressed in Section~\ref{DynamicNetworkWithTime-VaryingWeights}).

We would like to emphasize that, although the proposed dynamic network model generates zones and VNs in the same way as existing VN network models, the abstraction of dynamics is fundamentally different: the proposed model is a continuous-time model rather than a discrete-time model; the proposed model can represent the topology dynamics as well as the resource dynamics.

\subsection{Traffic Model}\label{TrafficModel}
For simplicity, the following traffic model is adopted in this paper without distinguishing applications. The task arrival is assumed to be Poisson stochastic processes with parameter $ \lambda $ because such processes have attractive theoretical properties~\cite{267444}. 
The $ k^{\rm th}\ (k\in Z^+) $ computational task arrives at VN $ u $ and instant $ t_{u,k} $ is denoted as $ \mathcal{T}_{u,k}$. It could be divided into multiple independent subtasks, i.e., $ \mathcal{T}_{u,k}= \{\tau_{u,k}^1,\tau_{u,k}^2,\dots,\tau_{u,k}^l,\dots,\tau_{u,k}^{n_{u,k}}\} \ (l,n_{u,k}\in Z^+,\ l\leq n_{u,k}) $, where $ \tau_{u,k}^l $ is the $ l^{\rm th}$ subtask of $ \mathcal{T}_{u,k}$ and $ n_{u,k} $ is the total number of subtasks of $ \mathcal{T}_{u,k}$. Subtasks are the smallest unit of transmission and computation. Subtasks belonging to the same task are routed to the same destination. 

Generally, there are seven items used to depict subtask $ \tau_{u,k}^l $, i.e., $ \Lambda_ {\tau_{u,k}^l} = (\widetilde{C_{u,k}^l},\widetilde{N_{u,k}^l},\widetilde{S_{u,k}^l},\vartheta_{u,k}^l, \alpha_{u,k}^l, \beta_{u,k}^l, t_{u,k}) $, where $ \widetilde{C_{u,k}^l} $, $ \widetilde{N_{u,k}^l} $, $\widetilde{S_{u,k}^l} $ and $ \vartheta_{u,k}^l $ are the computation requirement (i.e., necessary CPU cycles) of accomplishing subtask $ \tau_{u,k}^l $ in GFLOPS, the data volume of subtask $ \tau_{u,k}^l $ in gigabytes (GB), the amount of memory needed to complete the computation of subtask $ \tau_{u,k}^l $ in GB and the required delay threshold to process subtask $ \tau_{u,k}^l $ in seconds. $ \alpha_{u,k}^l$ and $\beta_{u,k}^l $ represent the longitude and latitude of the destination, respectively.

In addition, $ n_{u,k} $, $ \widetilde{C_{u,k}^l} $, $ \widetilde{N_{u,k}^l} $ and $ \widetilde{S_{u,k}^l} $ follow the log-normal distribution (i.e., 
$ \ln(n_{u,k})\sim N(\mu_n,\sigma_n^2) $, 
$ \ln(\widetilde{C_{u,k}^l})\sim N(\mu_C,\sigma_C^2) $, 
$ \ln(\widetilde{N_{u,k}^l})\sim N(\mu_N,\sigma_N^2) $ and 
$ \ln(\widetilde{S_{u,k}^l})\sim N(\mu_S,\sigma_S^2) $ ). The delay threshold of subtask $ \tau_{u,k}^l $ is randomly chosen from $ \{\vartheta_1,\vartheta_2\}$. The longitude $ \alpha_{u,k}^l $ and latitude $ \beta_{u,k}^l $ of the destination of $ \tau_{u,k}^l $ are randomly generated and subject to uniform distribution.

\subsection{Delay Model}\label{DelayModel}
In this paper, the overall delay consists of the transmission delay, the propagation delay, the computation delay and the waiting delay.

In dynamic networks, the \textit{transmission} delay $T_{trans}$ of subtask $ \tau_l $ on edge $e(i,j)$ starting from instant $ t $ satisfies the following equation: $\widetilde{N_l}=\int_{t}^{t+T_{trans}} R_{i,j}^r(t) dt$, where $\widetilde{N_l} $ is the data volume of subtask $\tau_l$ and $R_{i,j}^r(t)$ is the available transmission rate of edge $ e=(i,j) $ at instant $ t $.

Similarly, the \textit{computation} delay $T_{comp}$ of $ \tau_l $ at node $i$ starting from instant $t$ satisfies the following equation: $\widetilde{C_l}=\int_{t}^{t+T_{comp}}C_i^r(t) dt$, where $\widetilde{C_l}$ is the computational requirement of subtask $\tau_l$. $C_i^r(t)$ is the amount of available computing capability that node $ i $ can provide at instant $ t $.

Ignoring the minor distance changes during transmissions, the \textit{propagation} delay $T_{prop}$ on edge $e(i,j)$ starting from instant $ t $ is mathematically defined as $T_{prop}(t)=D_{i,j}(t)/c$, where $ D_{i,j}(t) $ represents the distance between node $ i $ and node $ j $ at instant $ t $. $ c $ is the speed of light. 

The \textit{waiting} delay is the duration from the arrival of a subtask to the moment when the subtask starts being transmitted or processed. In the following, the waiting delay before transmission and computation are included in the corresponding transmission delay and computation delay, respectively.

\section{Snapshot-Free Dynamic Network Modeling}\label{NetworkModeling}
To overcome the limitations of existing LEO satellite network models stated in Section~\ref{RelatedWork-Routing}, a snapshot-free dynamic network modeling method is proposed in this section. In this section, the resource dynamics are addressed first. Then we propose a graph-based method to address the topology dynamics. Finally, the advantages of the proposed network modeling method are summarized.

\subsection{Definition of Dynamic Network Model}
The proposed dynamic network model generates zones and VNs in the same way as existing VN network models. However, they have some significant differences: the proposed model is a continuous-time model rather than a discrete-time model; the proposed model can represent the topology dynamics as well as the resource dynamics. 

The snapshot-free dynamic network model (SFDNM) could be defined as $ G_{SFDNM}(t)=\big(\textbf{V},\textbf{E}(t),\textbf{W}_\textbf{E}(t),\textbf{W}_\textbf{V}(t)\big) $, where $ \textbf{V}=\{v_1,v_2,\dots,v_n\}(n=|\textbf{V}|)$ is the set of VNs, $ \textbf{E}(t)=\{e_1,e_2,\dots,e_{m(t)}\} (m(t)=|\textbf{E}(t)|)$ is the set of edges. An edge could be represented as $ e=(i,j) $ if $ i $ is the head of $ e $ and $ j $ is the tail of $ e $. $ \textbf{W}_\textbf{E}(t)=\{\omega_{e_1}(t),\omega_{e_2}(t),\dots,\omega_{e_{m(t)}}(t)\} $ is the weight set of edges and $ \textbf{W}_\textbf{V}(t) = \{\omega_{v_1}(t),\omega_{v_2}(t),\dots,\omega_{v_{n}}(t)\} $ is the weight set of node (i.e., VNs). Since both transmission and computing resources have impact on computing-aware routing problem stated in Section~\ref{Introduction}, the proposed model have both edge weights and node weights. These weights are task-related, which will be elaborated in Section~\ref{EdgeNodeWeightDefinitions}.

\subsection{Time-Varying Resources Modeling}\label{DynamicNetworkWithTime-VaryingWeights}
\subsubsection{Impact Factors of Edges and VNs}\label{ImpactFactorsOfEdgesAndVNs}
Due to the high-speed mobility and the limited on-board resources of LEO satellites, the routing process is impacted by many factors, including 1) the intermittent communications between LEO satellites, 2) the bandwidths of ISLs, and 3) the available on-board resources of computing, memory and energy. Among these factors mentioned above, factor 1) and factor 2) are related to the transmission process and form the impact factor set of edges, whereas factor 3) is about the computing process and forms the impact factor set of VNs. The definitions of the impact factor set of VNs edges are given below.

\textit{The impact factor set of edges $ \Lambda_{\textbf{E}}=\{\mathbb{D},\mathbb{B}\} $.} The intermittent communication between satellites is determined by the variation of distance caused by the relative motion between satellites. Here the inter-satellite distance matrix is represented as $ \mathbb{D}=\{D_{i,j}(t),\ i,j\in \textbf{V},\ t\in[0,T]\}$, where $D_{i,j}(t) $ is the distance between satellites associated with VN $ i $ and VN $ j $ at instant $ t $; $ T $ is the duration of simulation. $ \mathbb{B}=\{B_{i,j}^r(t),\ i,j\in \textbf{V},\ t\in[0,T]\} $ presents the available spectrum bandwidth matrix of inter-satellite links, where $ B_{i,j}^r(t) $ is the available spectrum bandwidth of the communication link between VN $ i $ and VN $j $ at instant $ t $. For inter-satellite links, $ R_{i,j}^r(t)= \sigma \times B_{i,j}^r(t) $ indicates that the data transmission rate of edge $ e=(i,j) $ at instant $ t $ is proportional to the spectrum bandwidth $ B_{i,j}^r(t) $.

\textit{The impact factor set of VNs $ \Lambda_{\textbf{V}}=\{\mathbb{C},\mathbb{S},\mathbb{E}\} $.} Here the available computing capability matrix is denoted as $ \mathbb{C}=\{C_i^r(t),\ i\in \textbf{V},\ t\in[0,T]\} $, the available memory matrix is defined as $ \mathbb{S}=\{S_i^r(t),\ i\in \textbf{V},\ t\in[0,T] \}$, and the available energy matrix is represented as $ \mathbb{E}=\{E_i^r(t),\ i\in \textbf{V},\ t\in[0,T] \}$. $ C_i^r(t) $, $ S_i^r(t) $, and $ E_i^r(t) $ represent the available computing capability in GFLOPS, memory resources in GB and battery energy in W$ \cdot $h of VN $ i $ at instant $ t $, respectively.

It is worth noting that the inter-satellite distance matrix $\mathbb{D} $ is derived from the satellite orbit parameters. In addition, the available bandwidth matrix $\mathbb{B} $, the available computing capability matrix $\mathbb{C}$, the available memory matrix $\mathbb{S} $, and the available energy matrix $\mathbb{E} $ are updated when the resources of the LEO satellite network change. Therefore, the proposed dynamic network model is update-driven and snapshot-free. 

\subsubsection{Edge Weights and Node Weights}\label{EdgeNodeWeightDefinitions}
For computing-aware routing, the routing process is composed of a transmission process and a computing process. That is, not only the data should be transmitted along the path, but also the tasks should be computed at the selected computing VNs on the path.
Specifically, the weight of an edge is related to the transmission process and is denoted as the sum of transmission delay and propagation\footnote{Propagation delays should be accounted for in LEO satellite networks, because the inter-satellite distance ranges from ten to thousands of kilometers. The propagation delays at milliseconds levels are much larger than those in terrestrial mobile networks~\cite{7466793}.} delay. The weight of a VN is related to the computing process and is defined as its processing delay. The equations for calculating the edge weights and node weights are defined as follows.

\textit{The set of edge weights.} $ \textbf{W}_\textbf{E}(t)=\{\omega_{e}^{u,k,l}(t) \mid e\in\textbf{E}(t),\ u\in\textbf{V},\ t\in[0,T]\}$ is the set of edge weights. $ \omega_{e}^{u,k,l}(t) $ is the weight of edge $ e=(i,j) $ corresponding to subtask $\tau_{u,k}^l$ and instant $ t $. It is the sum of the transmission delay $ T_{trans}^{e,u,k,l}(t) $ and propagation delay $ T_{prop}^{e}(t) $ of subtask $\tau_{u,k}^l$ on edge $ e=(i,j) $ starting from instant $ t $, which can be mathematically defined as follows.
\begin{equation}
		w_e^{u,k,l}(t)=
		\left\{
		\begin{array}{ll}
			T_{trans}^{e,u,k,l}(t)+T_{prop}^{e}(t),\ \zeta \leq \mathscr{T}_{i,j}^r(t), \\
			\infty,\ \mathrm{otherwise},
		\end{array}\label{Equ:1a}
		\right.
\end{equation}	

In equation~\eqref{Equ:1a}, $ \zeta=T_{trans}^{e,u,k,l}(t)+T_{prop}^{e}(t)$. $ \mathscr{T}_{i,j}^r(t) $ denotes the visible duration between $ i $ and $ j $ at $ t $ reflecting the intermittent communication of LEO satellite networks. It can be concluded that $ w_e^k(t) $ equals $ 0 $ when $ \zeta > \mathscr{T}_{i,j}^r(t) $, which indicates that task $ k $ cannot be transmitted through edge $ e=(i,j) $ successfully at $ t $ because the visible duration is shorter than the time required by the transmission process.

\textit{The set of node weights.} $ \textbf{W}_\textbf{V}(t) = \{\omega_{i}^{u,k,l}(t) \mid i,u\in\textbf{V},\ t\in[0,T]\} $ is the set of node weights. $ \omega_{i}^{u,k,l}(t) $ is the weight of VN $ i $ corresponding to subtask $\tau_{u,k}^l$ and instant $ t $. It is the processing delay $ T_{proc}^{i,u,k,l}(t) $ of $\tau_{u,k}^l$ at VN $ i $ start from instant $ t $, which can be mathematically defined as follows.
\begin{equation}
		w_i^{u,k,l}(t)=
		\left\{
		\begin{array}{ll}
			T_{proc}^{i,u,k,l}(t),\ S_i^r(t)\geq \widetilde{S_{u,k}^l}\ \& \ E_i^r(t)\geq f(\widetilde{C_{u,k}^l}) , \\
			\infty,\ \mathrm{otherwise},
		\end{array}\label{Equ:2a}
		\right.
\end{equation}	

In equation~\eqref{Equ:2a}, $S_i^r(t)$ and $E_i^r(t)$ are the amount of available memory and energy that VN $ i $ can provide at instant $ t $, respectively. $\widetilde{S_{u,k}^l}$ is the amount of memory required to complete subtask $ \tau_{u,k}^l$. $\widetilde{C_{u,k}^l}$ is the computation requirement of subtask $\tau_{u,k}^l$. $ f(\cdot) $ maps the amount of computation to the amount of energy consumption. It can be concluded that $ w_i^k(t) $ equals $ 0 $ when $ S_i^r(t) < \widetilde{S_{u,k}^l}$ or $E_i^r(t) < f(\widetilde{C_{u,k}^l})$, which means that the computation requirement of subtask $\tau_{u,k}^l$ cannot be completed by VN $ i $ at instant $ t $ if the VN's available memory or energy is insufficient.


Based on the statements above, it can be concluded that the snapshot-free dynamic network model has two main differences from the conventional static network models. First, the edges and nodes of the proposed dynamic network model are both weighted. For computing-aware routing, each subtask should be computed in the computing VN selected on the path. Since the processing delay is related to the available on-board resources of the selected computing VN, the nodes of the proposed dynamic network model should be weighted to assist the computing VN selection. Second, both the edge weights and the node weights are time-varying. Unlike static terrestrial networks, LEO satellites are moving at high speed. Therefore, what is changed is not limited to the network topology; the inter-satellite distances, the available ISL data rates, and the available on-board resources of satellites all change in real-time. The dynamics stated above are closely related to the routing path selection. It is necessary to model the LEO satellite network as a dynamic network with time-varying weights.

\subsection{Dynamic Topology Modeling}\label{Association}
\begin{figure}[t]
	\centering
	\setlength{\abovecaptionskip}{-0.cm}
	\setlength{\belowcaptionskip}{-0.cm}
	\includegraphics[width=0.48\textwidth]{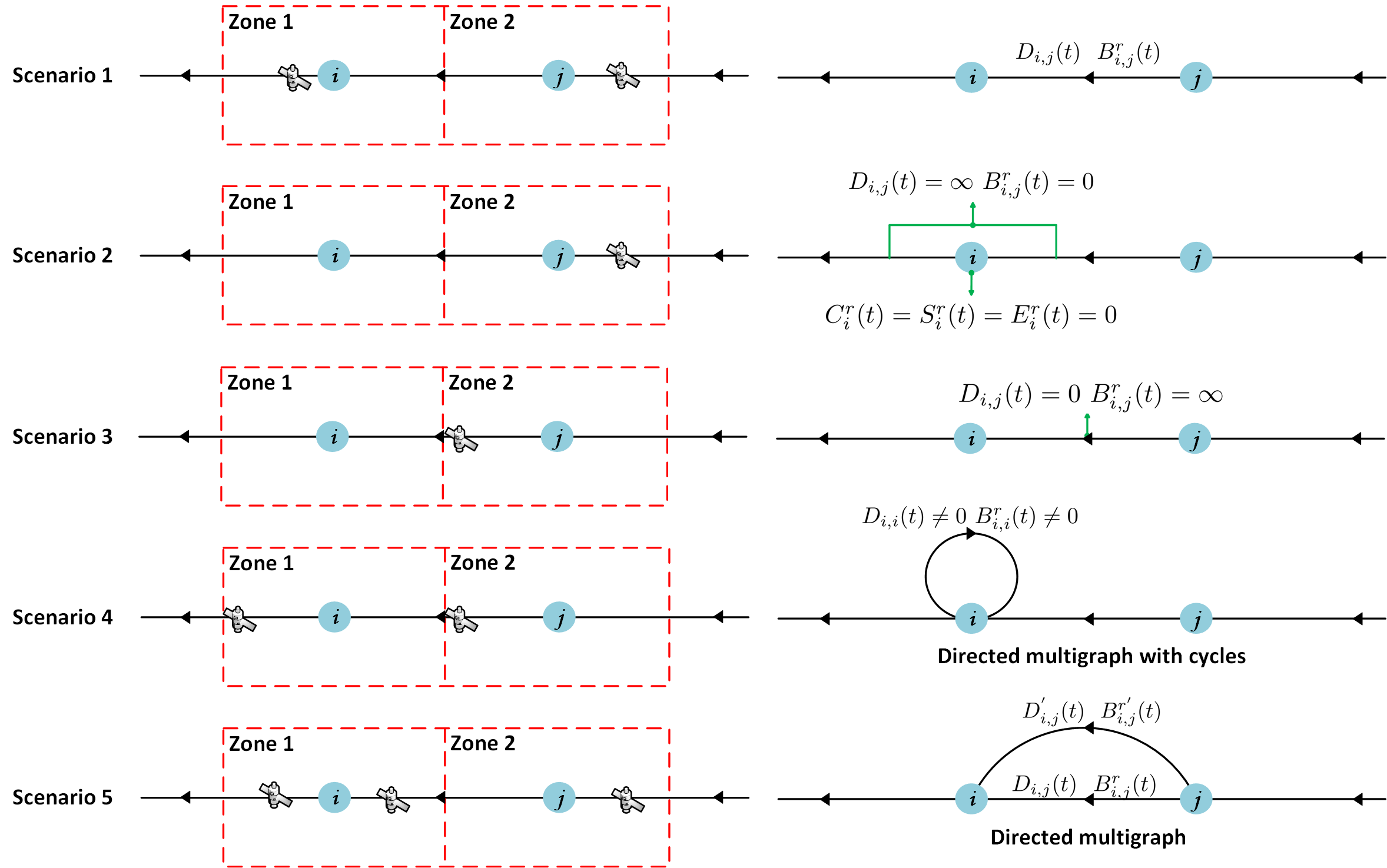}\\ 
	\begin{center}
		\caption{Association between satellites and VNs. The left side of the figure illustrates how the satellites move in each zone. The right side presents the corresponding changes in node weights and edge weights.}\label{ScenariosOfAssociationBetweenSatelliteAndVN}
	\end{center}
	\vspace{-1.0em}	 
\end{figure}

After modeling the LEO satellite network with the VN network model, the topology dynamics are converted into the dynamics of the association between satellites and VNs.
The numbers of satellites associated with two adjacent VNs and the positions of satellites located in the zones could lead to different types of edges between these two VNs. Since a complex scenario can be considered as a combination of several basic scenarios, as Fig.~\ref{ScenariosOfAssociationBetweenSatelliteAndVN} shows, we present five basic scenarios observed from the whole dynamic network, where two adjacent VN $ i $ and $ j $ are located in zone 1 and zone 2 respectively.

\textbf{Scenario 1:} at instant $ t $, both VN $ i $ and VN $ j $ are associated with a satellite and neither of the satellites has reached the boundaries of their corresponding zones, then there is an edge from $ j $ to $ i $ with attribute values $ D_{i,j}(t) $ and $ B_{i,j}^r(t) $.

\textbf{Scenario 2:} at instant $ t $, VN $ i $ is not associated with any satellite. In other words, there is no satellite in zone 1 and communications cannot be established between $ i $ and $ j $. In this case, the attribute values of the edge from $ j $ to $ i $ are $ B_{i,j}^r(t)=0 $ and $D_{i,j}(t)=\infty$.The available computing capability $ C_i^r(t) $, memory resources $ S_i^r(t) $ and energy storage $ E_i^r(t) $ of $ i $ all equal to $ 0 $. 

\textbf{Scenario 3:} at instant $ t $, VN $ i $ is not associated with any satellite; meanwhile, the satellite associated with VN $j $ just runs to the boundary between $ i $ and $j $, and a task needs to be transmitted from $j $ to $ i $. Although the task is transmitted from $j $ to $ i $, these two VNs are actually associated with the same satellite, thereby the inter-satellite transmission is not required. In this case, it can be considered that $D_{i,j}(t)=0$ and $ B_{i,j}^r(t)= \infty $. The transmission delay and the propagation delay of transmitting the task from $ j $ to $ i $ are both $ 0 $.

\textbf{Scenario 4:} at instant $ t $, satellite $ s_1 $ is associated with VN $ i $ and about to leave zone1. Satellite $ s_2 $ is about to enter zone 1. If a task computed by $ s_1 $ needs to be computed at VN $ i $ after instant $ t $, it needs to be transmitted from $ s_1 $ to $ s_2 $ at instant $ t $. During this process, although the satellites used for computing the task are changed, the VN for task execution remains the same. Therefore, VN $ i $ forms a loop at $t $: in this condition, tasks scheduled to be computed by $ i $ after $ t $ must be transmitted from VN $ i $ to itself through the loop to continue the execution.

\textbf{Scenario 5:} at instant $ t $, VN $ i $ is associated with multiple satellites. In this case, VN  $ i $ and its neighboring VN $ j $ are connected by multiple links, forming a multi-graph thereby. 

Based on the above five basic scenarios, the association between satellites and VNs at any moment and the switching of the association can be accurately represented graphically.

\section{Problem Formulation}\label{ProblemFormulation}
In this section, a typical DSSSP problem is introduced first, then the computing-aware routing problem is formulated and converted to a set of multiple DSSSP problems.


\subsection{Dynamic Single Source Shortest Path Problem}\label{DSSSP}
Many problems can be solved by searching the path with the minimum cost from the source node to the destination node. The cost model varies in different problems, for example, it can be time or the number of hops.
The problem becomes a DSSSP problem when the weight of each edge changes with the evolution of time~\cite{2000Fully,2001Semi,2000Maintaining}.

The DSSSP problems cannot be solved by traditional dynamic programming methods (such as the Dijkstra algorithm). It has aroused wide interests among researchers \comment{[1-17]}.

\textit{Problem $ \mathscr{P}0 $~\cite{Sunita2018Dynamizing}: }
Let $ G = \big(\textbf{V}, \textbf{E}(t), w(t)\big) $ be a simple directed graph, where $ \textbf{V}=\{V_1,V_2,\dots,V_n\}\ (n=|\textbf{V}|) $ and $ \textbf{E}(t)=\{e_1,e_2,\dots,e_{m(t)}\}\ (m(t)=|\textbf{E}(t)|) $ are the sets of vertices and edges, respectively.
Let $ e = (u, v) \in \textbf{E}(t) $; then $ u $ is the head of $ e $ denoted as $ e_h $, and $ v $ is the tail of $ e $ denoted as $ e_t $.
The edge weight function $ w(e,t) $ maps $ e \in \textbf{E}(t),\ t \in [0, T]$ to non-negative real numbers. It gives the weights of corresponding edges at instant $ t $\footnote{$ t $ is the instant when data is transmitted to the head of $ e $ (i.e., $ e_h $). It is also called ``the start time of $ w(e,t) $'', or ``the time of $ w(e,t) $'' for short. }.
In other words, the length of the path depends on time $ t $: 
assuming that there is a path $ P_{u,v}=\{(u_1,v_1),(u_2,v_2),\dots,(u_p,v_p)\}\ (u_1=u,\ v_p=v,\ u_p=v_{p-1},\ p\in Z^++1) $ and the start time is $ t_1 $,
then the length of path $ P_{u,v} $ is $ L_{t_1}^{P_{u,v}} =\psi(P_{u,v},t_1)= w(e_1,t_1)+w(e_2,t_2)+\cdots+w(e_p,t_p) $, where $ t_p=t_{p-1}+w(e_{p-1},t_{p-1})\ (p\in Z^++1) $, $ \psi(\cdot) $ maps the path and its start time to the path length.
Then the DSSSP problem is defined as finding the shortest path $ \pi_{u,v,t}=\phi(u,v,t) $ ($ \phi(\cdot) $ maps the source node, the destination node and the start time to the shortest path) and its length $ L_{t}^{\pi_{u,v}}$ from a specific source node $ u $ to each $ v\in \textbf{V} $ at time $ t $. 

It worth noting that $ \pi_{u,v,t}=\emptyset $ and $ L_{t}^{\pi_{u,v}}= \infty$ if $ v $ is not accessible from $ u $.
Problem $ \mathscr{P}0 $ is a non-convex optimization problem. It has been proved to be NP-hard~\cite{lin2007complexity}. In other words, it is computationally prohibitive to find an optimal solution directly for the optimization problem $ \mathscr{P}0 $. 

\subsection{Computing-Aware Routing Problem}\label{ComputingAwareRoutingProblem}
In this paper, subtasks are the smallest unit of transmission and computation; thus, subtasks are the unit of routing. Since subtasks cannot be further partitioned, there is only one computing node on each path in the proposed computing-aware routing scheme.
\subsubsection{Computing-Aware Routing Problem in $ G_{SFDNM}(t) $}\label{P1P2}
As stated above, the multipath-single-computing-node routing strategy is adopted in this paper. That is, multiple independent subtasks that make up a task could be routed (i.e., transmitted and computed) on different paths simultaneously. Each subtask is the smallest unit of transmission and computation and cannot be further partitioned.  

The ultimate goal of this paper is to find the optimal path for each subtask in the dynamic network $ G_{SFDNM}(t) $ established in Section~\ref{NetworkModeling} that could minimize the overall delay of each subtask, which could be formulated as follows.


\jiaqi{add rationale for optimization goals}
\textit{Problem $ \mathscr{P}1 $:} Let $ G_{SFDNM}(t)=\big(\textbf{V},\textbf{E}(t),\textbf{W}_\textbf{E}(t),\textbf{W}_\textbf{V}(t)\big)$ be a directed graph, where $ \textbf{V}=\{v_1,v_2,\dots,v_n\}\ (n=|\textbf{V}|)$ is the node set, $ \textbf{E}(t)=\{e_1,e_2,\dots,e_{m(t)}\}\ (m(t)=|\textbf{E}(t)|)$ is the edge set. 
Assuming that $ \mathcal{T}_{u,k} $ is the $ k^{\rm th}\ (k\in Z^+) $ computational task arrived at node $ u $ and its destination node is $ v $. For subtask $ \tau_{u,k}^l \in \mathcal{T}_{u,k}\ (l\in Z^+) $,  $ \textbf{W}_\textbf{E}(t)=\{\omega_{e}^{u,k,l}(t) \mid e\in\textbf{E}(t),\ u\in\textbf{V},\ t\in[0,T]\}$ is the set of edge weights and $ \textbf{W}_\textbf{V}(t) = \{\omega_{i}^{u,k,l}(t) \mid i,u\in\textbf{V},\ t\in[0,T]\} $ is the set of node weights. Assuming that subtask $ \tau_{u,k}^l $ is transmitted on path $P_{u,v}=\{(u_1,v_1),(u_2,v_2),\dots,(u_q,v_q),\dots,(u_p,v_p)\}\ (u_1=u,\ v_p=v,\ u_p=v_{p-1},\ p,q\in Z^+,\ p\geq q)$ and processed by $ u_q $ which is the selected computing node on path $ P_{u,v} $. If the start time of $ P_{u,v} $ is $ t_1 $, the length of path $ P_{u,v} $ is defined as $ L_{t_1}^{P_{u,v},u_q}=\Psi(P_{u,v},u_q,t_1)= \omega_{e_1}^{u,k,l}(t_1)+\omega_{e_2}^{u,k,l}(t_2)+\cdots+\omega_{e_{q-1}}^{u,k,l}(t_{q-1})+\omega_{u_q}^{u,k,l}(t_q)+\omega_{e_q}^{u,k,l}(t_q')+\omega_{e_{q+1}}^{u,k,l}(t_{q+1})+\cdots+\omega_{e_p}^{u,k,l}(t_p) $, where $ t_q $ is the instant that $ \tau_{u,k}^l $ arrives at node $ u_q $ (i.e., the head of edge $ e_q=(u_q,v_q) $), $ \Psi(\cdot) $ maps the path, the computing node and the start time to the path length. The relations of $ \{t_1,t_2,\dots,t_{q-1},t_q,t_q',t_{q+1},\dots,t_p\} $ are stated as follows,
\begin{subequations}
	\begin{align}
		t_2=&t_1+\omega_{e_1}^{u,k,l}(t_1), \label{Equ:3a}\\
		t_3=&t_2+\omega_{e_2}^{u,k,l}(t_2),\label{Equ:3b}\\ 
		&\dots \nonumber\\
		t_{q-1}=&t_{q-2}+\omega_{e_{q-2}}^{u,k,l}(t_{q-2}),\label{Equ:3c}\\ 
		t_q=&t_{q-1}+\omega_{e_{q-1}}^{u,k,l}(t_{q-1}), \label{Equ:3d}\\
		t_q'=&t_{q}+\omega_{u_q}^{u,k,l}(t_q),  \label{Equ:3e}\\
		t_{q+1}=&t_q'+\omega_{e_q}^{u,k,l}(t_q'),\label{Equ:3f}\\ 
		&\dots \nonumber\\
		t_p=&t_{p-1}+\omega_{e_{p-1}}^{u,k,l}(t_{p-1}). \label{Equ:3g}
	\end{align}
\end{subequations}
Then the computing-aware routing problem in $ G_{SFDNM}(t) $ (i.e., $ \mathscr{P}1 $) is defined as finding a path $ \pi_{u,v,t}$ from the source node $ u $ to the destination node $ v $ at time $ t $ and a computing node $ u_q $ on the path that can obtain the minimum path length $ L_t^{\Pi}\ (\Pi=\{\pi_{u,v,t},u_q\}=\Phi(u,v,t)$, $ \Phi(\cdot) $ maps the source node, the destination node and the start time to the shortest path) . 
Mathematically, the problem $ \mathscr{P}1 $ is formulated as 

\begin{subequations}
		\begin{align}
	(\mathscr{P}1):\ &
	\mathop{ \min}_{\Pi}\ (\Psi(\Pi,t)),\ \Pi=\{\pi_{u,v,t},u_q\}, \label{Equ:5a} \\
	{\rm s.t.}\ & u,v,u_q\in\textbf{V}, \label{Equ:5b}\\
          & t \geq 0. \label{Equ:5c}
		\end{align}
\end{subequations}

It could be concluded from the definitions stated above that there are some differences between $ \mathscr{P}0 $ and $ \mathscr{P}1 $. First, both the edges and nodes in $ \mathscr{P}1 $ are weighted. Second, besides the edges for transmission, in problem $ \mathscr{P}1 $, there is a computing node on the path to execute the computation of the corresponding subtask.

The shortest path of $ \mathscr{P}1 $ contains multiple transmission edges and a computing node. $ \mathscr{P}1 $ can be converted to $ \mathscr{P}0 $ when the computation is executed at the source node $ u $ or the destination node $ v $. Since $ \mathscr{P}0 $ is NP-hard, problem $ \mathscr{P}1 $ is NP-hard as well.

The computing-aware routing process in $ G_{SFDNM}(t) $ 
can be divided into three stages: (1) finding the shortest path from the source node $ u $ to the computing node $ u_q $ to transmit the raw data of subtask $ \tau_{u,k}^l $, (2) processing the computation of $ \tau_{u,k}^l $ on a specific computing node $ u_q $, (3) finding the shortest path from $ u_q $ to the destination node $ v $ to transmit the computation result of subtask $ \tau_{u,k}^l $. It is worth noting that stage (1) and stage (3) are both typical DSSSP problem (i.e., $ \mathscr{P}0 $) whose optimal results are $ \pi_{u,u_q,t}=\phi(u,u_q,t) $ and $ \pi_{u_q,v,t_q'}=\phi(u_q,v,t_q') $, respectively. Assuming that the path length of $ \pi_{u,u_q,t} $ and $ \pi_{u_q,v,t_q'} $ are $ L_t^{\pi_{u,u_q}}=\psi(\pi_{u,u_q,t},t) $ and $ L_{t_q'}^{\pi_{u_q,v}}=\psi(\pi_{u_q,v,t_q'},t_q') $. Then the problem $ \mathscr{P}1 $ can be rewritten as the following problem $ \mathscr{P}2 $.

\textit{Problem $ \mathscr{P}2 $:} Let $ G_{SFDNM}(t)=\big(\textbf{V},\textbf{E}(t),\textbf{W}_\textbf{E}(t),\textbf{W}_\textbf{V}(t)\big) $ be a directed graph. (The definition of $ G_{SFDNM}(t) $ here is the same as that in $ \mathscr{P}1 $.) For any subtask $ \tau_{u,k}^l \in \mathcal{T}_{u,k}\ (l\in Z^+) $, find a computing node $ u_q\ (u_q\in \textbf{V}) $, whose node weight is $ \omega(u_q,t_q)=\omega_{u_q}^{u,k,l}(t_q) $, which could minimize the path length $L_t^{\Pi}\ (\Pi=\{\pi_{u,v,t},u_q\}=\{\pi_{u,u_q,t}\cup\pi_{u_q,v,t_q'},u_q\} =\Phi(u,v,t))$. Mathematically, the problem $ \mathscr{P}2 $ is formulated as 
	
\begin{subequations}
		\begin{align}
 &(\mathscr{P}2):\nonumber\\
 &\mathop{ \min}_{u_q}\big(\psi(\phi(u,u_q,t),t)+\omega(u_q,t_q)+\psi(\phi(u_q,v,t_q'),t_q')\big), \label{Equ:7a}\\
 &{\rm s.t.}\  {\rm \eqref{Equ:5b} ,\ \eqref{Equ:5c}} \ and \nonumber \\ 
 &\quad\quad t\leq t_q <t_q'.\label{Equ:7b}
		\end{align}
\end{subequations}

For a specific subtask, $ u $, $ v $, $ t $ are known. The mapping of $ \psi(\cdot) $ is given in the definition of $ \mathscr{P}0 $. The mapping of $ \phi(\cdot) $ can be obtained by solving the DSSSP problem defined in $ \mathscr{P}0 $. In addition, $ t_q $ and $ t+q' $ can be calculated by \eqref{Equ:3d} and \eqref{Equ:3e}.

Given the above discussion, a computing-aware routing process can be separated into the transmission process (i.e., stage (1) and stage (3)) and the computing process (i.e., stage (2)). In this way, the problem $ \mathscr{P}1 $ can be converted to problem $ \mathscr{P}2 $ which divides the optimization procedure of $ \mathscr{P}1 $ into finding the shortest transmission path with a specific computing node and finding the optimal computing node with the corresponding shortest transmission path.

\section{Computing-Aware Routing Based on Snapshot-Free Dynamic Network Model}\label{AlgorihmDesign}
%
Section~\ref{NetworkModeling} provides an accurate model of the LEO satellite network studied in this paper. In this section, we discuss how to solve the computing-aware routing problem (presented in Section~\ref{P1P2}) in the established dynamic network model. 

Since the computing-aware routing problem in dynamic networks is NP-hard; thus, we propose the following GA-based algorithm to solve this problem as presented in Algorithm~\ref{algo:ga-routing}. 
In summary, it outputs the set of the optimal routing path and the computing node $ \Pi=\{\pi_t(u,v),u_q\} $ and the corresponding path length $ L^* $.

\begin{algorithm}[htbp]
	\caption{GA-based computing-aware routing\label{algo:ga-routing}}\small
	\KwData{The VN set and edge set of the dynamic network model $ \mathbf{V},\mathbf{E}(t) $}
	\KwData{Impact factor sets of edges and VNs \( \Lambda_{\mathbf{E}}=\{\mathbb{D},\mathbb{R}\} \),  $\Lambda_{\mathbf{V}}=\{\mathbb{C},\mathbb{S},\mathbb{E}\} $}
	\KwData{Subtask $ \tau_{u,k}^l $'s parameter set $\Lambda_ {\tau_{u,k}^l} = (\widetilde{C_{u,k}^l},\widetilde{N_{u,k}^l},\widetilde{S_{u,k}^l},\vartheta_{u,k}^l, \alpha_{u,k}^l,\beta_{u,k}^l, t_{u,k}) $
	}
	
	\KwResult{The optimal path and computing node set $ \Pi=\{P^*,u_q^*\} $ and the minimum overall delay $ L^* $}
	
	\KwResult{The updated \( \Lambda_{\mathbf{E}} \) and $\Lambda_{\mathbf{V}}$ }
	
	Initialize $ P^*\leftarrow\emptyset $, $ u_q^*\leftarrow\emptyset $, $ L^*\leftarrow\infty $, the instant set $ \mathbf{T}^*\leftarrow\emptyset $ \;
	
	Calculate $ \mathbf{W}_\mathbf{E}(t),\ \mathbf{W}_\mathbf{V}(t) $ based on $ \Lambda_{\mathbf{E}} $, $ \Lambda_{\mathbf{V}} $ and $ \Lambda_ {\tau_{u,k}^l} $ with Equ.~\eqref{Equ:1a} and~\eqref{Equ:2a};
	
	Set $ G_{SFDNM}(t) \leftarrow \big(\mathbf{V},\mathbf{E}(t),\mathbf{W}_\mathbf{E}(t),\mathbf{W}_\mathbf{V}(t)\big) $ \;	
	
	Set $\Lambda_ {\tau_{u,k}^l}^{'} = (\widetilde{C_{u,k}^l},0,\widetilde{S_{u,k}^l},\vartheta_{u,k}^l, \alpha_{u,k}^l,\beta_{u,k}^l, t_{u,k}) $ \;
	
	Calculate $ \mathbf{W}_\mathbf{E}^{'}(t) $ based on $ \Lambda_{\mathbf{E}} $ and $ \Lambda_ {\tau_{u,k}^l}^{'} $ with Equ.~\eqref{Equ:1a};
	
	Set $ G_{SFDNM}^{'}(t) \leftarrow \big(\mathbf{V},\mathbf{E}(t),\mathbf{W}_\mathbf{E}^{'}(t),\mathbf{W}_\mathbf{V}(t)\big) $ \;	
	
	\ForEach {$ u_q\in \mathbf{V} $} {
		\( (\pi_1, \mathbf{T}_1) \leftarrow GA(G_{SFDNM}(t), u, u_q, t_{u,k})\) \;
		\(d_1 \leftarrow L(\pi_1, t_{u, k})\) \;
		\(d_2 \leftarrow w_{u_q}^{u,k,l}(t_{u,k} + d_1) \) \;
		\((\pi_2, \mathbf{T}_2) \leftarrow GA(G_{SFDNM}^{'}(t), u_q, v, t_{u,k} + d_1 + d_2) \) \;
		\(d_3 \leftarrow L(\pi_2, t_{u, k} + d_1 + d_2)\) \;
		$ L_{temp} \leftarrow d_1 + d_2 + d_3 $ \;
		\If{$ L_{temp}<L^* $} {
			$ \Pi = \{P^*,\ u_q^*\} \leftarrow \{\pi_1 \cup \pi_2,u_q\}$ \;
			$ L^* \leftarrow L_{temp} $ \;
			$ \pi_1^* \leftarrow \pi_1$ \;
			$ \mathbf{T}^* \leftarrow \mathbf{T}_1  $	\;
			$ T_{proces}^* \leftarrow d_2  $ \;
		}
	}
	
	\For {$ i \leftarrow 1,2,\dots,|\pi_1^*|-1 $} {
		$R_{i,j}^r([\mathbf{T}^*(i),\mathbf{T}^*(i+1)]) \leftarrow 0 $ \;
	}
	$ C_{u_q^*}^r([\mathbf{T}^*(|\pi_1^*|),\mathbf{T}^*(|\pi_1^*|)+T_{proces}^*) \leftarrow 0 $ \;	
\end{algorithm}

As Algorithm~\ref{algo:ga-routing} shows, before running the algorithm, the sets of VNs and edges should be generated based on the dynamic network model construct method introduced in Section II first. In addition, the impact factor set of edges $ \Lambda_{\mathbf{E}} $, the impact factor set of VNs $ \Lambda_{\mathbf{V}} $ and subtask $ \tau_{u,k}^l $'s parameter set $ \Lambda_ {\tau_{u,k}^l} $ are the input data. 
Algorithm~\ref{algo:ga-routing} outputs the set of the optimal routing path and the computing node $ \Pi=\{\pi_t(u,v),u_q\} $, the corresponding minimum overall delay $ L^* $, the updated impact factor set of edges $ \Lambda_{\mathbf{E}} $ and the updated impact factor set of VNs $\Lambda_{\mathbf{V}}$.

Except the initialization (line 1), Algorithm~\ref{algo:ga-routing} contains three parts. The first part is the graph generation (line 2 to line 6). The algorithm prepares the key data structure \(G_{SFDNM}\) and assigns weight to it according to physical constraints. Inside the algorithm, the sets of edge weights and node weights of the dynamic network model $ G_{SFDNM}(t) $ are calculated based on $ \Lambda_{\mathbf{E}} $, $ \Lambda_{\mathbf{V}} $ and $ \Lambda_ {\tau_{u,k}^l} $ with Equ.~\eqref{Equ:1a} and~\eqref{Equ:2a} (line 2 and line 3).
Because the data volume of $ \tau_{u,k}^l $ can be reduced to a few bits after being computed, this paper only considers the propagation delay and ignores the transmission delay in the computing result transmission process. Therefore, a new edge weight set is calculated and a new dynamic network model $ G_{SFDNM}^{'}(t) $ is generated (line 4 to line 6).

The second part is the optimal path and computing node finding.
The main loop (line 7) iterates through all nodes. In each iteration, \(u_q\) is selected\footnote{Note that the complexity of the proposed computing-aware routing scheme can be significantly reduced if the computing nodes are selected in an appropriate range and order. This paper mainly focuses on evaluating how much overall delay can be reduced by the proposed computing-aware routing scheme compared with the ground-offloading approach. The computing node selection strategy will be investigated in-depth in the following research.} for computing the subtask, and the overall delay under this scenario is evaluated as \(L_{temp}\) which has three components (i.e., \(d_1\), \(d_2\) and \(d_3\)). 
\(d_1\) is caused by the transmission of raw data from the source node \(u\) to the computing node \(u_q\). The time at which the subtask is created is \(t_{u,k}\). The shortest path \(\pi_1\) and the instance set $ \mathbf{T}_1 $ (recording the instances when $ t_{u,k} $ first arrived at each node on $ \pi_1 $) are first solved with GA (line 8), and the delay \(d_1\) is calculated thereafter (line 9).
\(d_2\) is caused by the processing of the subtask. Since the subtask is arrived at \(t_{u,k} + d_1\) which affects the computation delay, the delay is computed as \(w_{u_q}^{u,k,l}(t_{u,k} + d_1)\) (see line 10).
\(d_3\) is caused by the computing result transmission from the computing VN \(u_q\) back to the destination VN \(v\). Similar to the calculation of the first part of the delay, the shortest path is first solved with GA (line 11) and the delay is calculated next (line 12). In this step, the transmission delay of each edge is ignored because the computation results are usually only a few bits; therefore, $ G_{SFDNM}^{'}(t) $ rather than $ G_{SFDNM}(t) $ is adopted as the input of GA.
After calculating the delay under the assumption that node \(u_q\) is selected for computing, the global state is updated to find the best node. Here the optimal path \(P^*\), the optimal computing node \(u_q^*\) and the minimum overall delay \(L^*\) are updated in line 15 and 16, respectively.
Furthermore, for updating $ \Lambda_{\mathbf{E}} $ and $ \Lambda_{\mathbf{V}} $ in the next part, the optimal path from $ u $ to $ u_q^* $, the instance set $ \mathbf{T} $ and the processing delay of the optimal computing node are recorded in line 17--19. 

The third part is impact factor sets update.
After the second part, the route for current subtask is resolved as $ \Pi \leftarrow \{P^*,u_q^*\} $. However, the offloading of the subtasks occupies computation and transmission resources of the network. To reflect these changes, line 23 to line 25 subtracts the resources occupied by $ \Pi $ from the impact factor values of the time when $ \tau_{u,k}^l $ arrives at $ u_q $ and each edge on $ \pi $.

It is worth noting that the proposed computing-aware scheme and the ground-offloading scheme are not mutually exclusive, but complementary. If computing resources on satellites are not sufficient, tasks can be offloaded to ground servers for computing.  By adding ground servers as nodes of the dynamic network model, these ground servers will be selected for computing if the delay of on-board computing is larger than the delay of ground offloading.

\section{Simulation Results and Analyses}\label{Simulations} 
In this section, we evaluate the GA-based computing-aware routing scheme proposed in Section~\ref{AlgorihmDesign}, and analyze the simulation results.
We aim to answer the following research questions:
\begin{itemize}
	\item \textbf{RQ1}: How much computing capability is required for LEO satellites to enable the proposed scheme to effectively reduce delay? 
	\item \textbf{RQ2}: How does the transmission capability affect the performance of the proposed scheme?
	\item \textbf{RQ3}: What kind of tasks can be accelerated by the proposed scheme?
\end{itemize}

Furthermore, we would like to emphasize that the following assumptions and settings are made in the simulations.
\begin{itemize}
	\item \emph{Sufficient energy and storage resources.} As shown in Table~\ref{Onboard Computing Systems}, existing onboard computing systems can provide thousands of GFLOPS of computing capability. Therefore, we assume that modern LEO satellites can afford the energy consumption when the computation capability is less or equal than 400 GFLOPS. The detailed influence of energy consumption will be discussed and evaluated in future work.
	\item \emph{Zero computation delay for ground-offloading routing scheme.} Ground servers (such as supercomputers) usually have large amounts of computational resources. We round the computation delay towards zero for a fair comparison.
	\item \emph{Platform-agnostic evaluation.} A satellite platform contains multiple components. For example, the onboard computing system determines the computing capability, whereas the signal transmitters and receivers determine the transmission capability. Since the components can be composed at will, a variety of configurations can be found for LEO satellites. Rather than evaluating specific LEO satellite configurations exhaustively, we reveal how the factors affect the performance in general.
	\item \emph{Scheme-level comparison.} Since the pathfinding algorithm adopted in the ground-offloading scheme can also be applied to the transmission process of the proposed computing-aware scheme, for fairness, the pathfinding algorithm for both schemes is set as the GA algorithm. In other words, we focus on scheme-level rather than algorithm-level comparison in this paper.
	Furthermore, since task partitioning strategy also affects the routing performance, for a fair comparison, subtask is assumed to be the smallest unit of transmission and computation in both schemes.
\end{itemize}

The simulation parameters are presented in Table~\ref{SimulationParameters}.
\begin{table}[h]
	\centering
	\begin{spacing}{1}
		\caption{Simulation Parameters}\footnote{The maximum transmission rate reflects the transmission capacity of the satellite. In actual transmissions, the transmission rate is proportional to the bandwidth, which decreases as resources are occupied and increases as resources are released (never greater than the maximum value). In this way, the available resources are time-varying. Accordingly, the weights of the proposed model are time-varying. Similarly, the available computing resources are time-varying.}\label{SimulationParameters}
	\end{spacing}
	\scriptsize
	\begin{tabular}{cc}
		\toprule
		Parameter & Value\\
		\midrule
		Radius of the earth $ R_e $                  & 6,371,393 m\\
		Mass of the earth $ M_e $                    & $ 5.965\times10^{24}$ kg\\
		Earth rotation angular velocity $ \omega_e $ & $ 7.29211510\times 10^{-5} $ \\
		Gravitational  $ G $                         & $ 6.67428\times 10^{-11} $\\
		Kepler constant $ K $                        & $ 3.9860\times 10^{14} $\\
		Velocity of light $ c $                      & $ 299,792,458$ m/s\\	
		Number of orbits                             & 10\\
		Satellites per orbit                         & [10,12,10,12,10,\\
		\                                            & 12,10,12,10,12]\\        
		Orbit altitude $ h $                         & [200,300,400,500,600,\\
		\                                            & 200,300,400,500,600] km\\
		Orbit inclination $ i_0 $                    & $ 90^{\circ} $\\
		Max data rate of ISL                         & 5 Gbps\\
		Channel number per ISL                       & 1 \\
		Max data rate of SGL      & 0.2 Gbps\\	
		Max computing capability     &  \\    
		of satellites $ C $     & 100 GFLOPS\\ 
		The number of tasks arrived    & \ \\
		at each VN per second $ \lambda $    & 1/60\\
		Distribution parameters of  & \ \\
		subtask number per task $ (\mu_n,\sigma_n) $& (3,1)\\
		Distribution parameters of subtask's & \ \\
		computation requirement $ (\mu_C,\sigma_C) $ & (50,2) GFLO\\
		Distribution parameters of  & \ \\	
		sub-task's data volume $ (\mu_N,\sigma_N)  $   & (0.1,0.02) Gbps\\
		\bottomrule
	\end{tabular}
\vspace{-1.0em}
\end{table}

\subsection{Reduced Overall Delay (RQ1)}
\begin{figure}[htbp]
	\centering
	\setlength{\abovecaptionskip}{-0.cm}
	\setlength{\belowcaptionskip}{-0.cm}
	\includegraphics[width=0.5\textwidth]{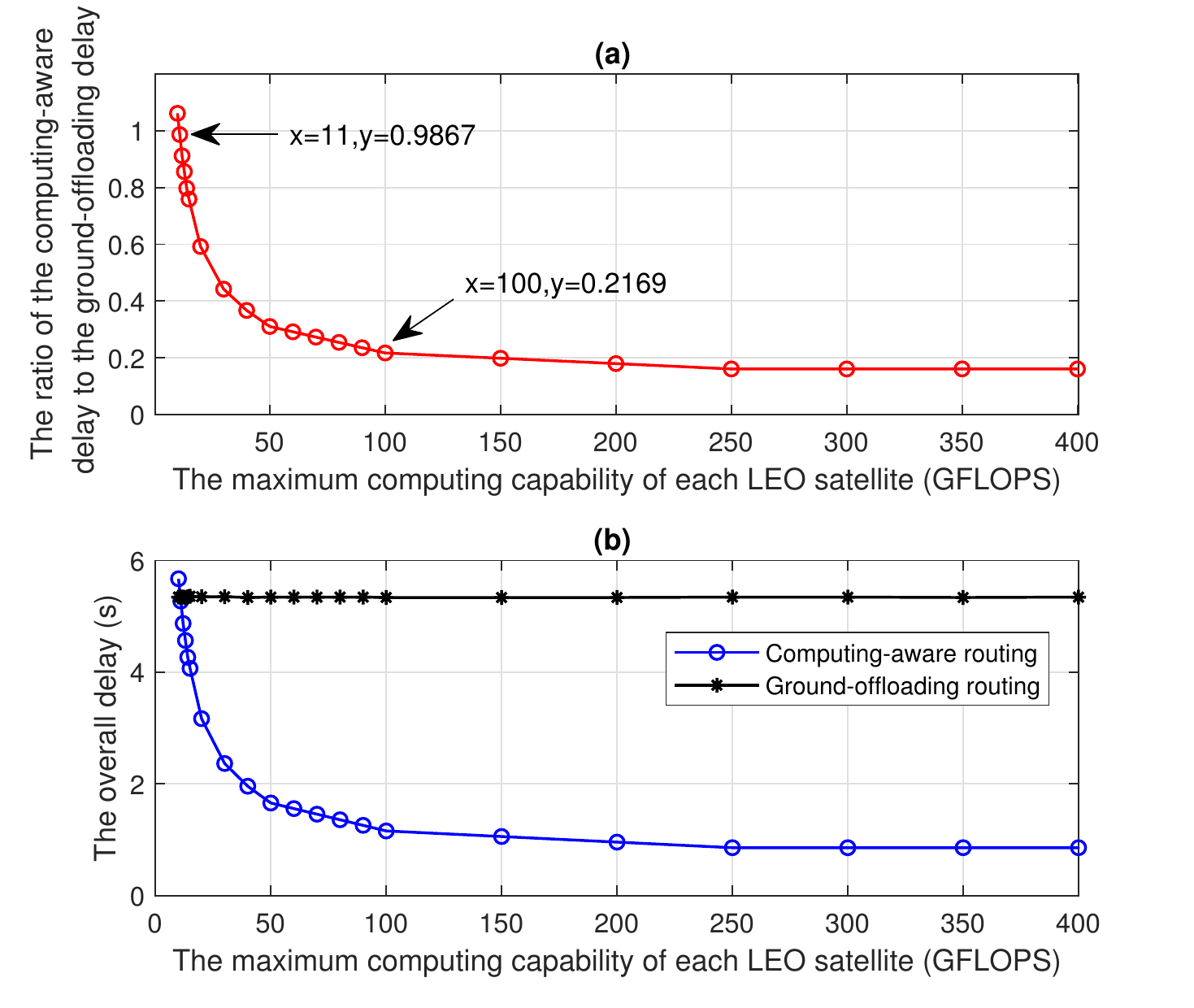}\\
	\begin{center}
		\caption{(a) The ratio of the computing-aware delay to the ground-offloading delay versus the maximum computing capability of each LEO satellite $ C $. A ratio of 1.0 means that two methods have the same performance. The lower the ratio, the better the proposed method. (b) The overall delays of the computing-aware routing scheme and the ground-offloading routing scheme}\label{SimFig1}
	\end{center}	 
	\vspace{-1.0em}
\end{figure}

\begin{figure}[htbp]
	\centering
	\setlength{\abovecaptionskip}{-0.cm}
	\setlength{\belowcaptionskip}{-0.cm}
	\includegraphics[width=0.5\textwidth]{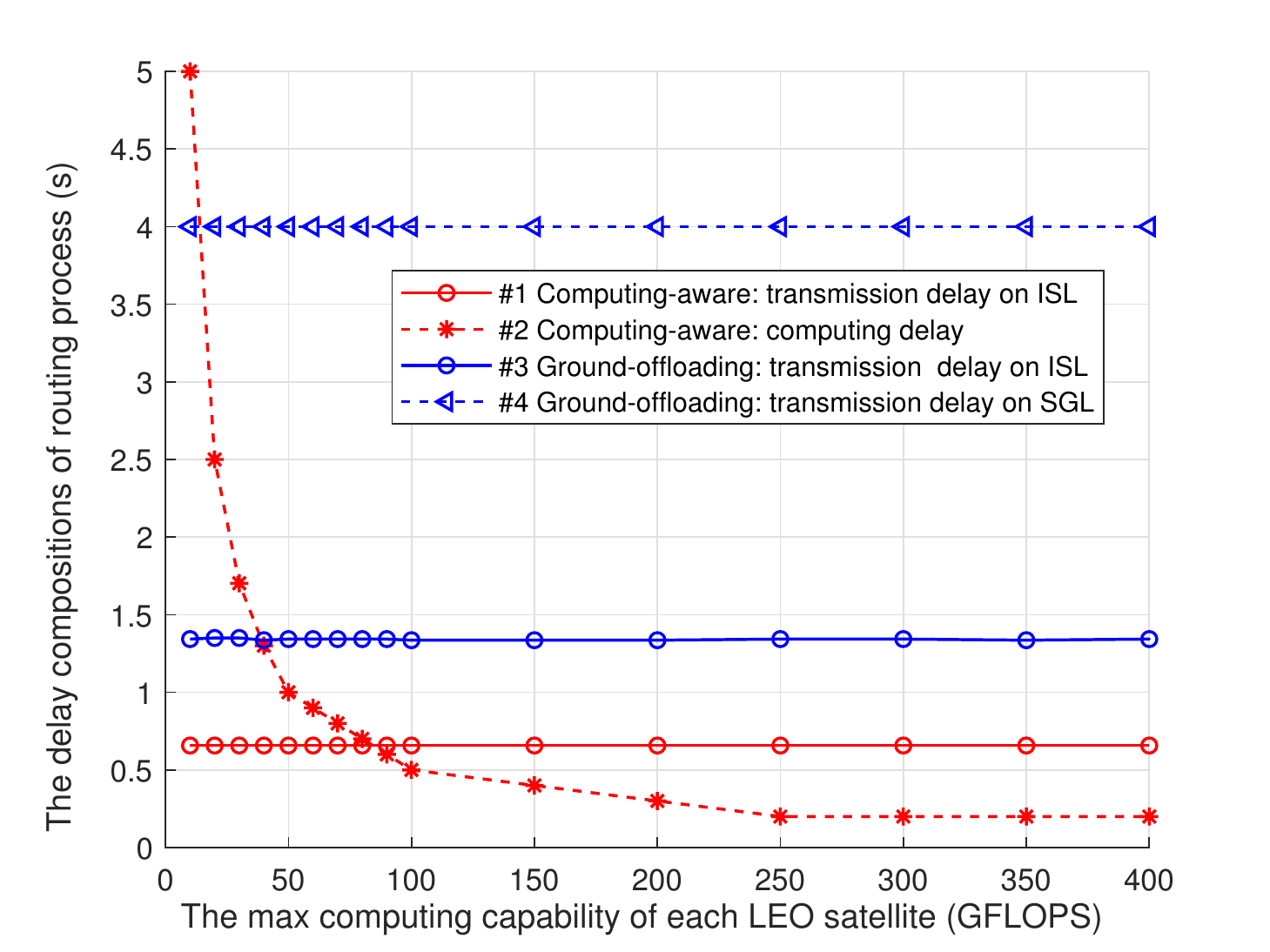}\\
	\begin{center}
		\caption{Delay decomposition of computing-aware routing scheme v.s. ground-offloading routing scheme. The overall delay of computing-aware routing  mainly consists of the transmission delay on ISL (\#1), and the processing delay (\#2). Similarly, the overall delay of ground-offloading routing mainly consists of the transmission delay on ISL (\#3), and the transmission delay on SGL (\#4).}\label{SimFig2}
	\end{center}
\end{figure}

Fig.~\ref{SimFig1} and Fig.~\ref{SimFig2} shows how the computing capability of LEO satellites affects the performance of computing-aware routing schemes. The simulation covers computing capability from 100 GFLOPS to 400 GFLOPS, and the remaining parameters follow Table~\ref{SimulationParameters}. For each computing capability, the delay of the computing-aware routing scheme is calculated, and the result is normalized with the corresponding delay of the benchmark routing scheme.

Fig.~\ref{SimFig1} (a) demonstrates shortened delays from the proposed computing-aware routing scheme. 
It can be concluded that the proposed computing-aware routing scheme can reduce the overall delay at most computing capabilities ($ C > 11 $ GFLOPS). 
Especially when the maximum computing capability is set to 100 GFLOPS, the overall delay is reduced to 21.69\% compared to the benchmark scheme (i.e., 3.61$\times$ speedup). 
Since existing onboard computing systems can already provide these computing capabilities, the proposed scheme is not only effective but also feasible.

Furthermore, we notice that when $ C < 11 $ GFLOPS, the performance of the ground-offloading scheme is better, which indicates that the time saved by reducing the amount of data transmitted is less than the increased computation delay due to insufficient computing capability.
In this condition, tasks can be offloaded to ground servers for computing. In other words, the proposed computing-aware scheme and the ground-offloading scheme are not mutually exclusive, but complementary. By adding ground servers as nodes of the dynamic network model, these ground servers will be selected for computing if the delay of on-board computing is larger than the delay of ground-offloading.

In addition, the curve in Fig.~\ref{SimFig1} (a) shows that the stronger the computing capability, the greater the reduction. It flattens out as more computing capability is added to satellites. This is because the computation delay of the proposed scheme converges to 0 after the computing capability is increased to a certain value. Meanwhile, the other part (i.e., transmission delay) of the overall delay remains constant. In this way, the overall delay of the proposed scheme converges to a constant value and the curve flattens out.
The diminishing return indicates the cost-effectiveness of the computing-aware routing scheme. The most cost-saving hardware configuration is allocated to each satellite with 100 GFLOPS of computing capability. It is less than 10\% of the most powerful existing onboard computing systems listed in Table~\ref{Onboard Computing Systems}.
In other words, not only can computing-aware routing achieve significant performance boost with advanced hardware, the advantage also persists with budget hardware with minimum costs.

The reasons for the change in ratio in Fig.~\ref{SimFig1} (a) can be explained by Figure~\ref{SimFig1} (b): as the computing capability of LEO satellites grows, the proposed approach can leverage more onboard resources to complete the tasks without transmitting them back to the ground. Therefore, the delay of the computing-aware routing gradually decreases, while the delay of ground-offloading routing stays still.

To further explain why computing-aware routing decreases the delay, Fig.~\ref{SimFig2} presents a delay decomposition of both the computing-aware routing scheme and the ground-offloading routing scheme.

Fig.~\ref{SimFig2} shows that the processing delay of the computing-aware routing scheme decreases when the computing capability of LEO satellites increases, whereas other delays remain unchanged.
Furthermore, it can be concluded that the transmission delay of the computing-aware routing scheme is much lower than that of the benchmark scheme. It demonstrates that the proposed computing-aware routing scheme decreases delay successfully by transmitting computation results instead of raw data. When the computing capability is greater than 11 GFLOPS, the delay reduced by transmitting computing results is larger than the increased computation delay; the ratio in Fig.~\ref{SimFig1} decreases when the computing capability of LEO satellites increases.

\begin{mdframed}
 	The proposed scheme can accelerate the offloading with a computing capability as low as 11 GFLOPS under the provided settings.
 	When the computing capability is increased to 100 GFLOPS, a trivial computing capability supported by most LEO satellites, a 3.61$\times$ speedup can be observed.
	In general, the reduction becomes more significant as computing capability grows.
\end{mdframed}

\subsection{Impact from Transmission Capability (RQ2)}
\begin{figure}[htbp]
	\centering
	\setlength{\abovecaptionskip}{-0.cm}
	\setlength{\belowcaptionskip}{-0.cm}
	\includegraphics[width=0.5\textwidth]{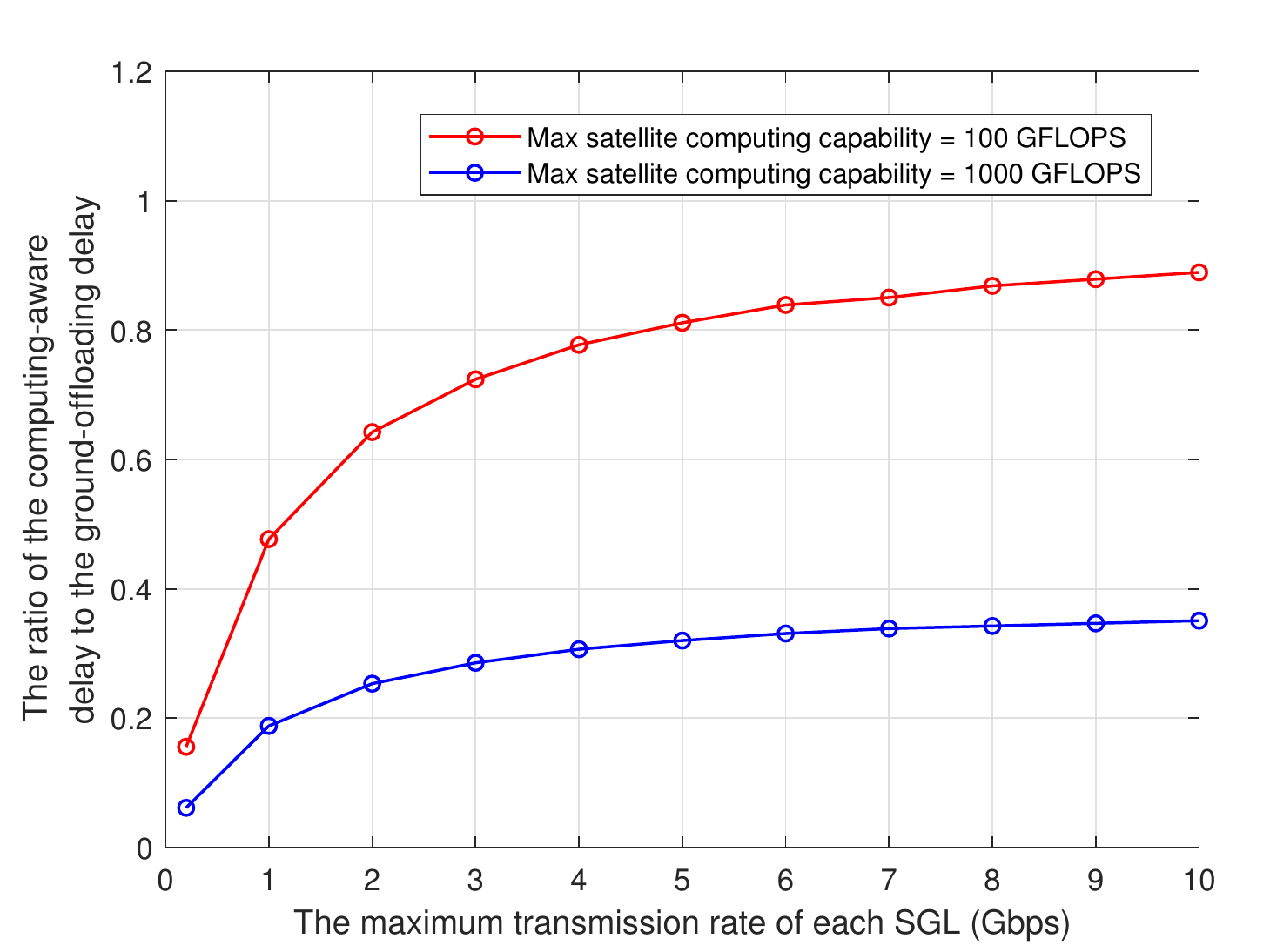}\\
	\begin{center}
		\caption{Delay of the computing-aware routing scheme under different SGL transmission rates, normalized to the ground-offloading scheme.}\label{SimFig5}
	\end{center}
\end{figure}

\begin{figure}[htbp]
	\centering
	\setlength{\abovecaptionskip}{-0.cm}
	\setlength{\belowcaptionskip}{-0.cm}
	\includegraphics[width=0.5\textwidth]{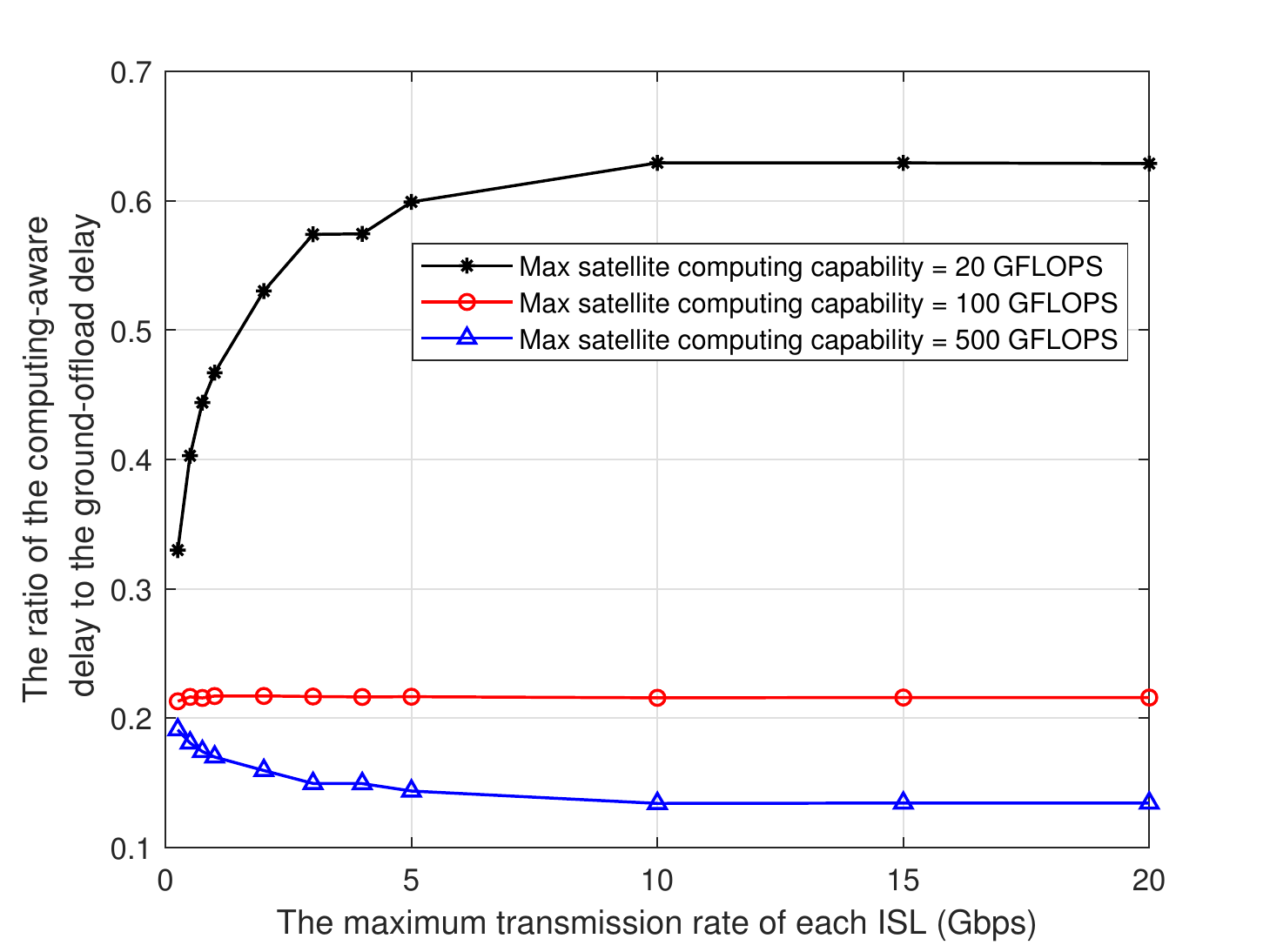}\\
	\begin{center}
		\caption{Delay of computing-aware routing under different ISL transmission rates, normalized to the ground-offloading delay.}\label{SimFig8}
	\end{center}
\end{figure}

Fig.~\ref{SimFig5} and Fig.~\ref{SimFig8} show how the transmission rates of SGL and ISL affect the performance of the computing-aware routing scheme. The simulation covers transmission rate of SGL from 0.2 Gbps to 10 Gbps and transmission rate of ISL from 0.25 Gbps to 20 Gbps. The remaining parameters follow Table~\ref{SimulationParameters}.

In Fig.~\ref{SimFig5}, the leftmost points presents the situation of the most common cases at the moment, where the transmission rate of SGL is assumed to be 0.2 Gbps. Similarly, the rightmost points present the ideal scenarios where the transmission rate of SGL is set to 10 Gbps.
It could be concluded that, with the most practical SGL transmission rate setting at present, i.e., 0.2 Gbps, satellites with 100 GFLOPS and 1000 GFLOPS of computing capability can save 84.43\% and 93.86\% of the overall delay, respectively. 
Moreover, in Fig.~\ref{SimFig5}, the ratio increases with the increase of SGL transmission rate. This is because  the ground-offloading scheme transmits a much larger amount of data over the SGL than the proposed computing-aware scheme; therefore, as the SGL transmission capacity increases, the reduction in transmission delay over the SGL is much greater for the ground-offloading scheme than for the proposed computing-aware scheme.
Although the advantage of computing-aware routing scheme diminishes with a higher quality SGL, it is still significant: when the bandwidth is boosted to 10 Gbps, the satellite with 100 GFLOPS achieves a speedup of 11.07\% nevertheless.

Fig.~\ref{SimFig8} shows the performance at different ISL rates. The ratio can be approximated with $ (T_{comp}+x\times \Delta_{ISL})/(T_{trans,SGL}+y\times \Delta_{ISL}) $, where $ T_{comp} $ is the computation delay of the proposed scheme, $ T_{trans,SGL} $ is the transmission delay on SGL of the ground-offloading scheme. $ \Delta_{ISL} $ is the average delay on each ISL (i.e., one hop). $ x $ and $ y $ are the average hop number of transmitting raw data on ISL for the proposed scheme and the benchmark scheme, respectively. Since the proposed scheme performs onboard computing, $ x<y $. After deducing the above equations, the following conclusion can be obtained: when $ T_{comp}/T_{trans,SGL}=x/y $, the ratio constantly equals $ x/y $; when $ T_{comp}/T_{trans,SGL}>x/y $, the ratio is larger than $ x/y $ and vice versa. In addition, when increasing the transmission capability of ISL, $ \Delta_{ISL} $ will decrease and converge to $ 0 $; therefore, the three curves all converge to $ T_{comp}/T_{trans,SGL} $. With this conclusion, we can determine whether to perform the proposed scheme (onboard computing) or perform the ground-offloading scheme instead by estimating $ T_{comp}/T_{trans,SGL} $.

\begin{mdframed}
	The proposed scheme outperforms the ground-offloading scheme for all the realistic ISL/SGL configurations used in simulations.
	For the worst case of SGL ($C = 20$ GFLOPS, $R_{ISL} = 20$ Gbps), the proposed scheme still reduces 37.12\% of the baseline delay.
	For the worst case of ISL ($C = 100$ GFLOPS, $R_{SGL} = 10$ Gbps), the proposed scheme still reduces 11.07\% of the baseline delay.
\end{mdframed}

\subsection{Impact from Task Properties (RQ3)}

\begin{figure}[htbp]
	\centering
	\setlength{\abovecaptionskip}{-0.cm}
	\setlength{\belowcaptionskip}{-0.cm}
	\includegraphics[width=0.5\textwidth]{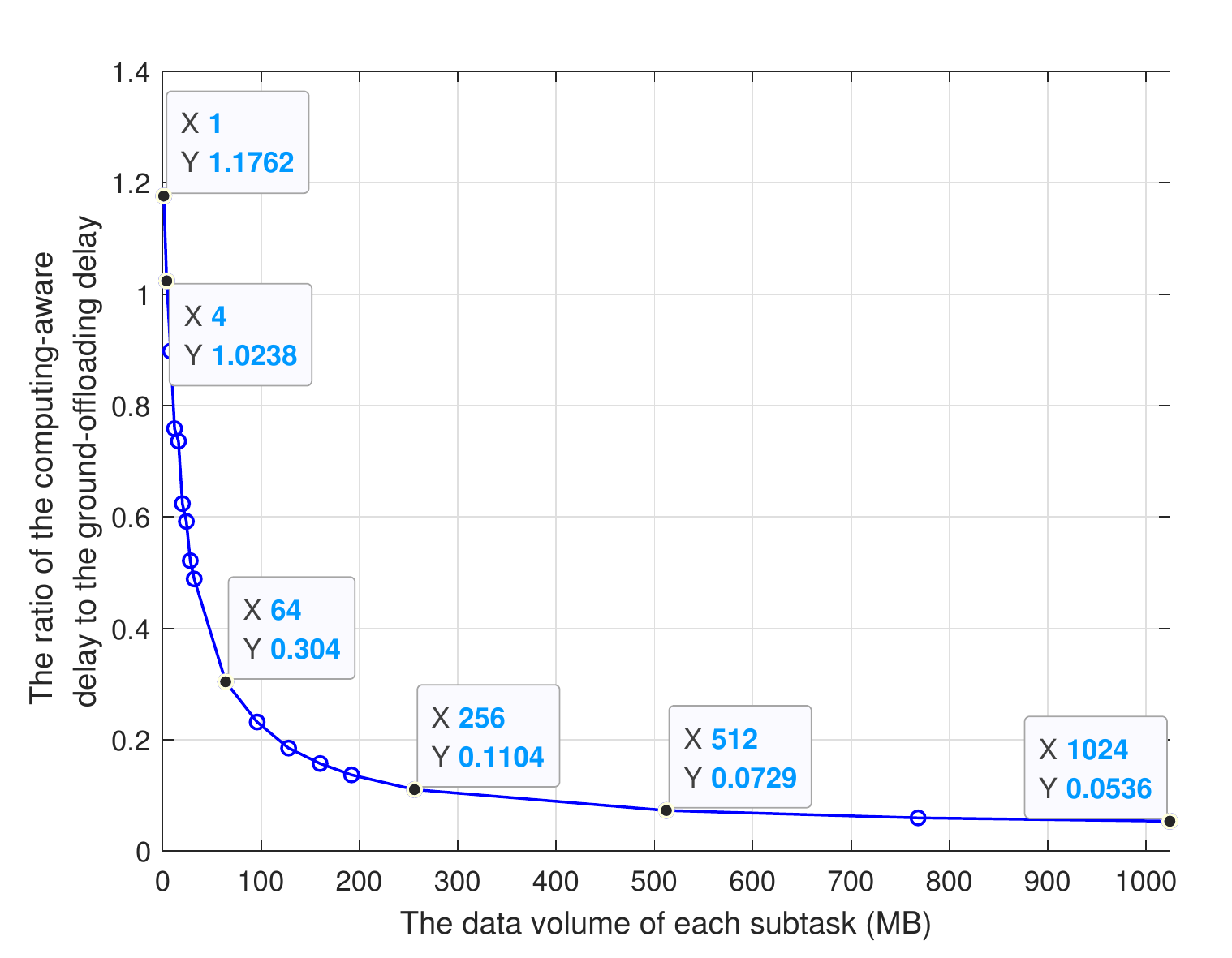}\\
	\begin{center}
		\caption{Performance impact from subtasks' data volumes. The delay of the computing-aware routing scheme is normalized with the delay of the ground-offloading scheme. In other words, a ratio lower than 1.0 means that the proposed method is better.}\label{SimFig3}
	\end{center}
\end{figure}
\begin{figure}[htbp]
	\centering
	\setlength{\abovecaptionskip}{-0.cm}
	\setlength{\belowcaptionskip}{-0.cm}
	\includegraphics[width=0.5\textwidth]{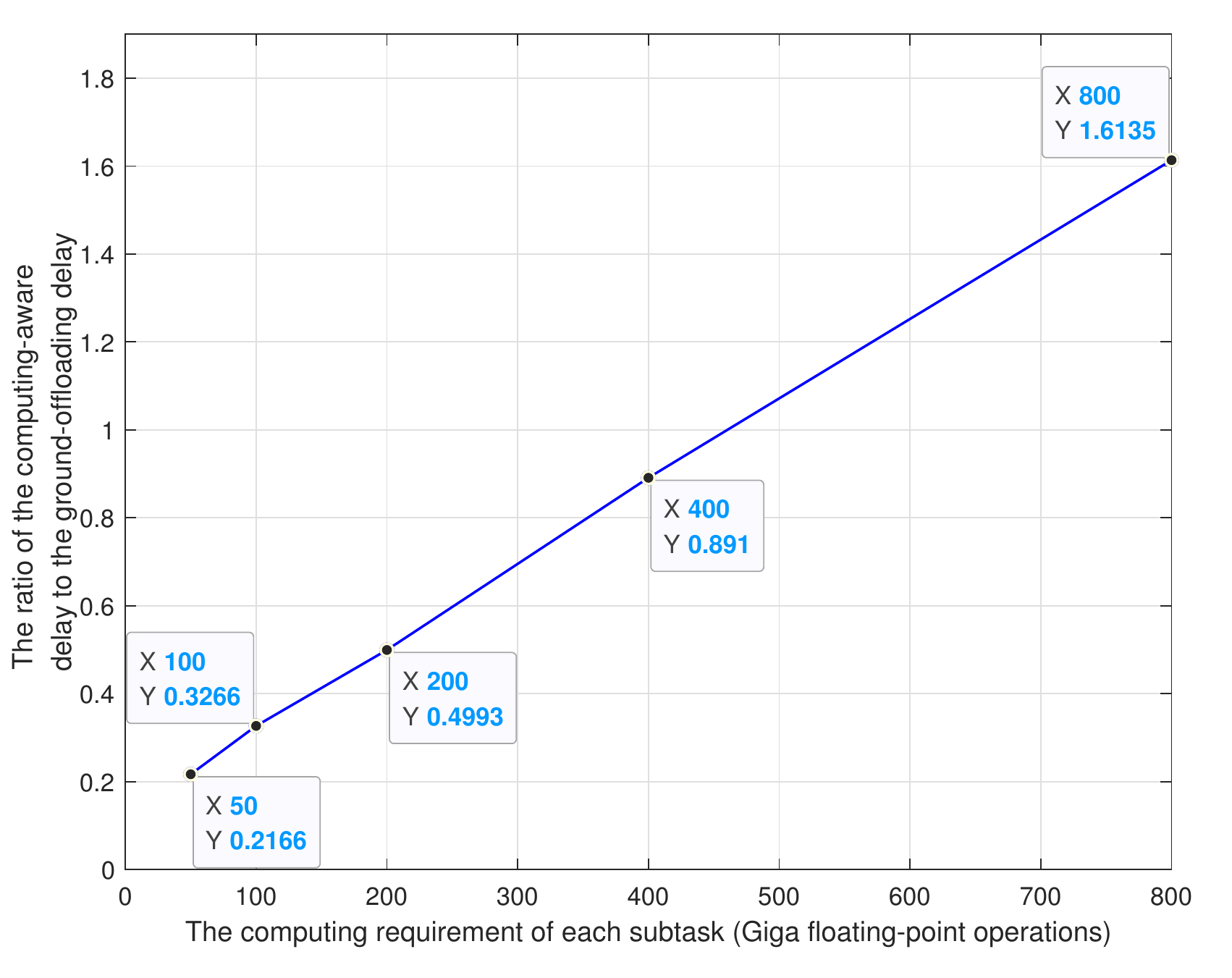}\\
	\begin{center}
		\caption{Performance impact from subtasks' computing requirements. the delay of the computing-aware routing is normalized with the ground-offloading delay. In other words, a ratio lower than 1.0 means that the proposed method is better.}\label{SimFig4}
	\end{center}
\end{figure}
\begin{figure}[htbp]
	\centering
	\setlength{\abovecaptionskip}{-0.cm}
	\setlength{\belowcaptionskip}{-0.cm}
	\includegraphics[width=0.5\textwidth]{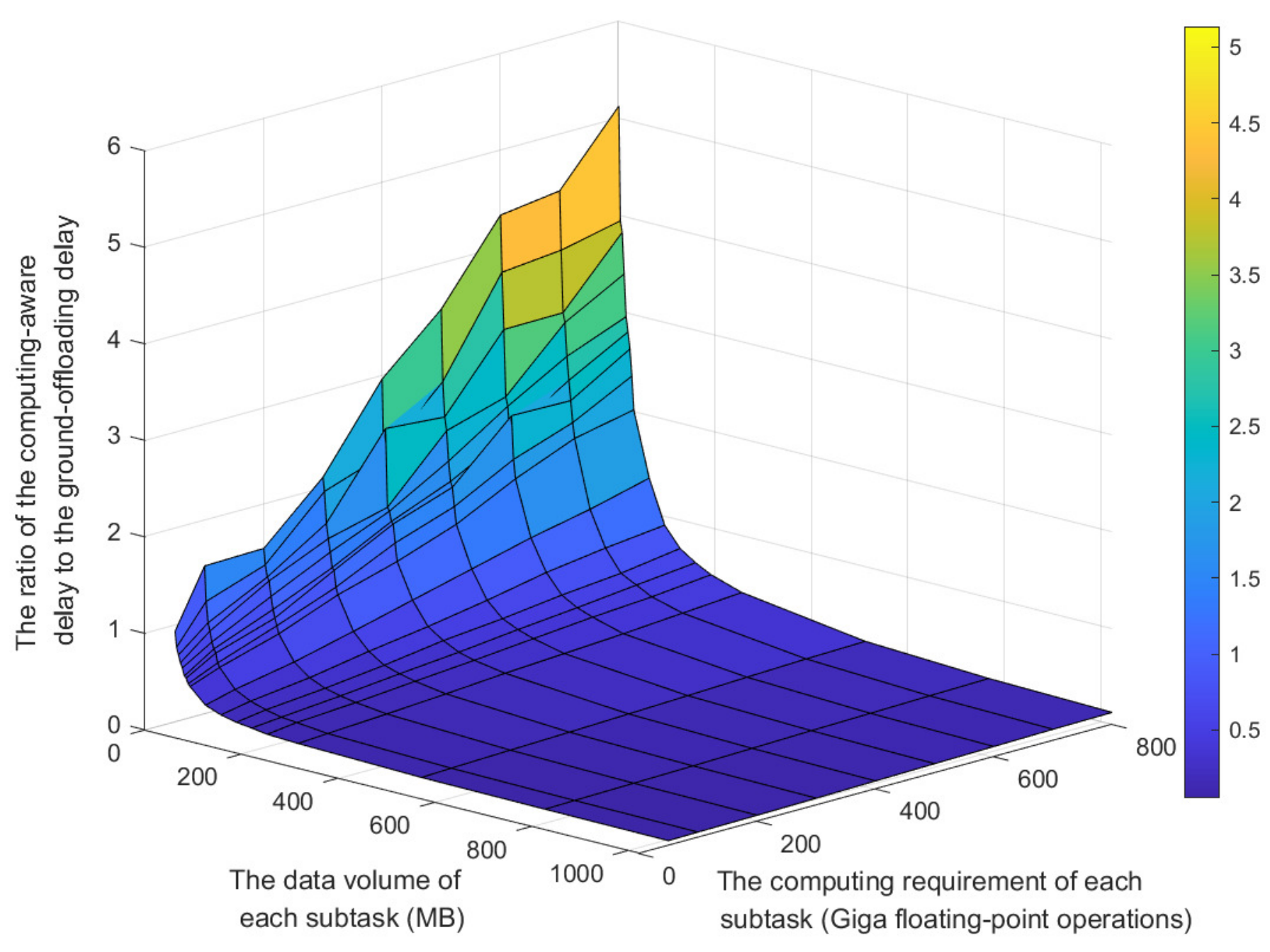}\\
	\begin{center}
		\caption{The impact of subtasks' data volume and computing requirements on the delay of the computing-aware routing scheme (normalized to the ground-offloading delay). A ratio of 1.0 means two methods have equally good performance. The lower the ratio, the better the proposed method is.}\label{SimFig7}
	\end{center}
\end{figure}

Fig.~\ref{SimFig3}, Fig.~\ref{SimFig4}, and Fig.~\ref{SimFig7} show how the data volume and computing requirement of subtasks affects the performance of the computing-aware routing scheme, respectively. The simulation covers data volumes from 1 MB to 1 GB, and computing requirements from 50 Giga floating-point operations (GFLO) to 800 GFLO. The remaining parameters follow Table~\ref{SimulationParameters}. In Fig.~\ref{SimFig3}, Fig.~\ref{SimFig4}, and Fig.~\ref{SimFig7}, the delay of the computing-aware routing is calculated, and the result is normalized with the corresponding delay of ground-offloading.

From Figure~\ref{SimFig3} we can conclude that the proposed computing-aware routing scheme could reduce the overall delay over a wide range of data volume ($>4$ MB) with current network settings. The superiority of the computing-aware routing becomes more and more significant when the data volume of each subtask increases. Since the computing-aware routing replaces transmissions of large-scale raw data with final results, the amount of data to transfer is greatly reduced. 
The reduced demand also opens up opportunities in future data-intensive applications: for example, for applications transferring 1GB of data, the task execution efficiency can be boosted by 17.66x with computing-aware routing.

Figure~\ref{SimFig4} shows how the computing requirement of subtasks affects the performance of the proposed computing-aware routing. The simulation covers computing requirements from 50 GFLO to 800 GFLO, and the remaining parameters follow Table~\ref{SimulationParameters}.

From Figure~\ref{SimFig4} we can conclude that the computing-aware routing scheme has linear scalability concerning the computing requirement of each subtask. Because the onboard computing resources are limited in nature, computing-aware routing is not intuitively suitable for tasks with extreme computing demands. Despite this, for most tasks (400 GFLO or less), the proposed computing-aware routing scheme still handles them well; for corner cases where computing-aware routing is not the best approach, its performance downgrades gracefully in a linear way as the computing requirements grow.

Figure~\ref{SimFig7} presents a comprehensive view of the delay of the computing-aware routing scheme, encompassing both the variables of subtasks' data volumes and computing requirements. It provides a method for compensating the limitations of computing-aware routing on tasks with high computation costs. Specifically, the strategy to determine whether to use the computing-aware routing  scheme or fallback to the conventional ground-offloading routing for a specific task can be pre-computed by the ground station or GEO satellites.

When the deployment of a satellite network has completed, the parameters of onboard resources (e.g. computing capability) and the data rate of ISLs and SGLs are known in advance. Therefore, the ground station can predict the overall performance of computing-aware routing for a specific task by considering its computing requirement and data volume. The evaluation results (as Figure~\ref{SimFig7} shows) can be uploaded to the satellite network in advance, and all newly generated tasks follow the predetermined threshold accordingly.

\begin{mdframed}
	The proposed scheme is applicable to most kinds of tasks. Unless the task has minimal data volume (less than 4MB) or extremely high computation requirements (more than 400 GFLO), the proposed scheme can be used for improving offloading performance.
\end{mdframed}

\section{Conclusion}\label{Conclusions}
This paper investigates the LEO satellite network routing to fulfill new requirements of space missions. It first analyzes the challenges in LEO satellite networks, including the highly dynamic network topology, limited onboard resources, and intensive computational demands.

Aiming at tackling the challenges, this paper proposes a computing-aware routing scheme for LEO satellite networks.
The paper first models the dynamic set of satellites as a snapshot-free network with time-varying weights. Then the computing-aware routing problem in the dynamic network is formulated as a combination of multiple DSSSP problem. In addition, a GA-based method is proposed to approximate the results in reasonable time. Simulation results demonstrate the applicability of the proposed approach, where the overall delay can be reduced in a wide range of network settings.

In the future, we will study how to further optimize the proposed computing-aware routing scheme to reduce its complexity. Furthermore, we will work on the task splitting mechanism in computing-aware routing.
\bibliographystyle{IEEEtran}
\bibliography{IEEEfull,Reference}

\end{document}